\documentclass[twocolumn,preprintnumbers,amsmath,amssymb,superscriptaddress,nofootinbib,longbibliography, prl]{revtex4-2}

\usepackage[T1]{fontenc}
\usepackage[utf8]{inputenc}
\usepackage{lmodern}
\usepackage{amsmath,amssymb,mathtools}
\usepackage{comment}
\usepackage{physics}
\usepackage{bm}
\usepackage{microtype}
\usepackage{xcolor}
\usepackage{overpic}
\usepackage[normalem]{ulem}
\usepackage{soul}
\usepackage{hyperref}
\hypersetup{colorlinks=true,linktoc=all,linkcolor=blue,breaklinks=true,citecolor=blue,urlcolor=blue}
\usepackage{orcidlink}
\usepackage{cleveref}
\Crefname{equation}{Eq.}{Eqs.}
\Crefname{figure}{Fig.}{Figs.}
\Crefname{section}{Sec.}{Secs.}

\newcommand{\subfigref}[2]{\ref{#1}\hyperref[#1]{(#2)}}

\begin{document}

%\title{Thermodynamic uncertainty relation and minimum Fano factor of a coherent thermoelectric heat engine}
\title{Thermodynamic bound on the Fano factor of a coherent thermoelectric heat engine}

\author{Nahual Sobrino~\orcidlink{0000-0001-5704-693X}}
\affiliation{The Abdus Salam International Centre for Theoretical Physics, Strada Costiera 11, 34151 Trieste, Italy}

\author{Rosario Fazio~\orcidlink{0000-0002-7793-179X}}
\affiliation{The Abdus Salam International Centre for Theoretical Physics, Strada Costiera 11, 34151 Trieste, Italy}
\affiliation{Dipartimento di Fisica, Universit\`a di Napoli ``Federico II”, Monte S. Angelo, I-80126 Napoli, Italy }
\author{Matteo Acciai~\orcidlink{0000-0001-9739-9289
}}
\affiliation{The Abdus Salam International Centre for Theoretical Physics, Strada Costiera 11, 34151 Trieste, Italy}

\begin{abstract}
    We show that for fermionic coherent thermoelectric transport, selecting heat-engine operation yields a thermodynamic uncertainty relation for the charge current
    that imposes a universal lower limit of $F > 1/2$ on the corresponding Fano factor. We find that violations of the thermodynamic uncertainty relation for classical Markov processes, typically associated with a quantum advantage, are far more restricted for heat engines than what is allowed by a generic thermodynamic process. 
    For bosonic and classical carriers, the minimum Fano factor increases to $F > 1$, and the thermodynamic uncertainty relation for classical Markov processes is never violated. We provide numerical evidence that all the obtained bounds are tight and can be saturated by properly designed transmissions.
\end{abstract}
\maketitle

%%%%%%%%%%%%%%%%%%%%%%%%%%%%%%%%%%%%%%%%%%%%%%%%%%%%%%%%%%%%%%%%%%%%%%%%%%%%%%%%%%%%%%%%%%%%%%%%%%%%%%%%%%%%%%%%%%%%%%%%%%%%%%%%%%%%%%%%%%%%

The study of current fluctuations in quantum transport has a long history~\cite{Blanter2000Sep}, and their importance is effectively summarized in the famous quote ``the noise is the signal'' by R. Landauer~\cite{Landauer1998Apr}, highlighting noise as a useful source of information to characterize mesoscopic conductors.
In recent years, there has been a fruitful cross
fertilization of ideas coming from quantum transport and stochastic thermodynamics~\cite{Strasberg2024Aug}.
A particularly relevant example of interplay between these two fields is the research on Thermodynamic Uncertainty Relations (TURs), one of the most active lines of modern thermodynamics~\cite{Binder2018,Campbell2026Jan}. First derived for classical, Markovian systems~\cite{Barato2015Apr,Gingrich2016Mar}, the prototypical TUR, which will be henceforth called \emph{classical TUR}, bounds the precision of a generic (average) current $\mathcal{J}$ with an entropic cost in the form $S_\mathcal{J}\sigma/\mathcal{J}^2\ge 2k_\mathrm{B}$, where $\sigma$ is the rate of entropy production and $S_\mathcal{J}$ is the current fluctuation. This means that driving a current with little fluctuations, or high precision, comes at the cost of a large dissipation. These findings initiated a quest to find more refined thermodynamic bounds. Considerable effort was put in generalizing the classical TUR in a variety of contexts, and considering a \emph{generic thermodynamic process} in the same spirit of the original result. This led to several extensions both in the classical realm~\cite{Polettini2016Nov,Brandner2018Mar,Proesmans2019May,Dechant2020Mar,Falasco2020May,Vo2022Sep,Koyuk2022Nov,Dechant2026May} as well as in quantum systems~\cite{Macieszczak2018Sep,Agarwalla2018october,Saryal2019october,Guarnieri2019Oct,Timpanaro2019Aug,Falasco2020Sep,Hasegawa2021Jan,Potanina2021Apr,VanVu2023Feb,Salazar2024Jan,Hegde2026Mar,Vidal2026May}, including hybrid superconducting devices~\cite{Mayo2026Mar,Sobrino2026May,Ohnmacht2025Dec,Vidal2026Jun}, and feedback processes~\cite{Potts2019Nov,Honma2026Feb}.
In particular, for coherent transport, several bounds on fluctuations were reported~\cite{Eriksson2021Sep,Acciai2024Feb,Tesser2024May,Tesser2026Mar,Pan2026Jan}, and quantum TURs constraining the precision of the charge current for fermionic systems~\cite{brandner2025thermodynamic,Brandner2025Oct} and (very recently) generic currents~\cite{Tesser2026Jul} were derived.

\begin{figure}[t]
	\centering
	\includegraphics[width=\linewidth]{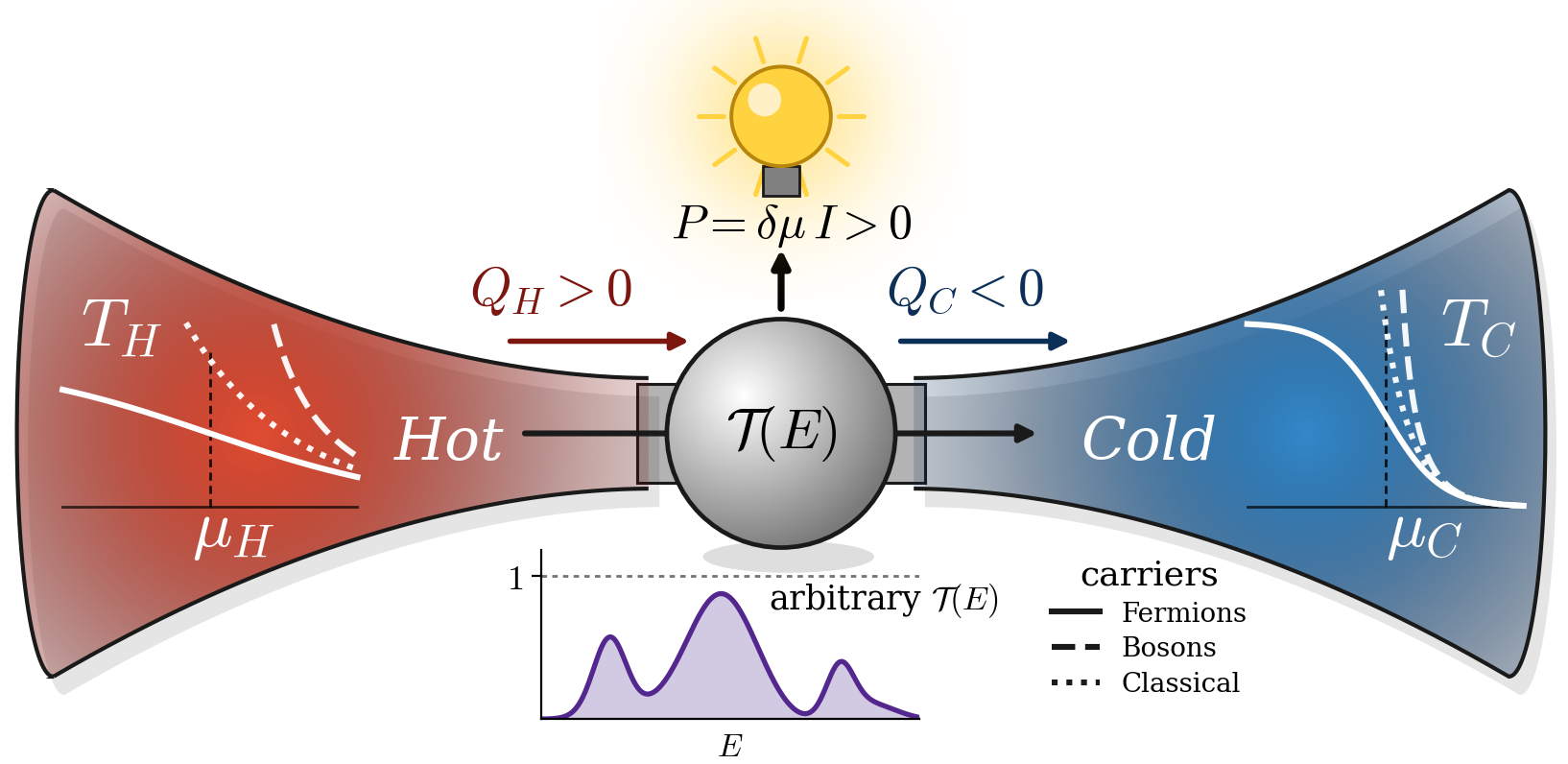}
	\caption{Sketch of a coherent thermoelectric heat engine operating between a hot ($T_H,\mu_H$) and a cold ($T_C,\mu_C$) reservoir of fermionic, bosonic, or classical carriers, connected through a scattering region. %If the coherence length of the carriers exceeds the system size and no inelastic collisions occur,
    Transport properties are encoded in a transmission function $\mathcal{T}(E)$.}
	\label{fig1}
\end{figure}

Soon after its discovery, it was realized that the TUR can be used to estimate the efficiency $\eta$ of steady-state heat engines (HEs) operating between temperatures $T_H$ and $T_C$ and producing power $P=\delta\mu\, I$, where $I$ is the charge current (see Fig.~\ref{fig1}). Indeed, a direct rewriting yields in this case~\cite{Pietzonka2018May}
\begin{equation}
    \frac{\eta}{\eta_\mathrm{c}}\le\left[1+\frac{k_{\mathrm{B}}T_C}{|\delta\mu|} \frac{2}{F}\right]^{-1}\,,
\label{eq:efficiency_TUR_cl}
\end{equation}
where $F=S_I/|I|$ is the Fano factor, $\delta\mu=\mu_C-\mu_H$ the difference of chemical potentials,  and $\eta_\mathrm{c}$ Carnot's efficiency.
%%%%%%%%%%%%%%%%%%%%%%%%%%%%%%%%%%%%%%%%
%%%%%%%%%%%%%%%%%%%%%%%%%%%%%%%%%%%%%%%%
This result shows that $\eta=\eta_\mathrm{c}$ can be reached either at zero power or diverging fluctuations, as $F\to\infty$ in either case. On the contrary, $\eta\to 0$ is obtained as $F\to 0$.
However, since the classical TUR is not specifically tailored to describe a heat engine, but rather gives a constraint valid for an arbitrary thermodynamic process, there is no guarantee that~\eqref{eq:efficiency_TUR_cl} is a tight bound. Therefore, some works have investigated whether and how Eq.~\eqref{eq:efficiency_TUR_cl} is modified when targeting a heat engine.
For instance,
Ref.~\cite{Kamijima2021Oct} found a more restrictive bound that involves higher-order cumulants of $I$, which are shown to be relevant in the nonlinear regime. Even in linear response, constraint~\eqref{eq:efficiency_TUR_cl} can be tightened by introducing the figure of merit $ZT$ for a heat engine~\cite{Kheradsoud2019Aug}.
The case of \emph{cyclic thermal machines} was addressed in~\cite{Koyuk2019June,Miller2021May,Eglinton2022May,Lu2022Mar,VanVu2024Apr}. In particular, Ref.~\cite{Miller2021May} showed that, thanks to the cyclic operation, it is possible to achieve a higher efficiency than what is allowed by~\eqref{eq:efficiency_TUR_cl}, but quantum fluctuations are always detrimental.

%%%%%%%%%%%%%%%%%%%%%%%%%%%%%%%%%%%%%%%%
\begin{figure}[t]
	\centering
	\includegraphics[width=1\linewidth]{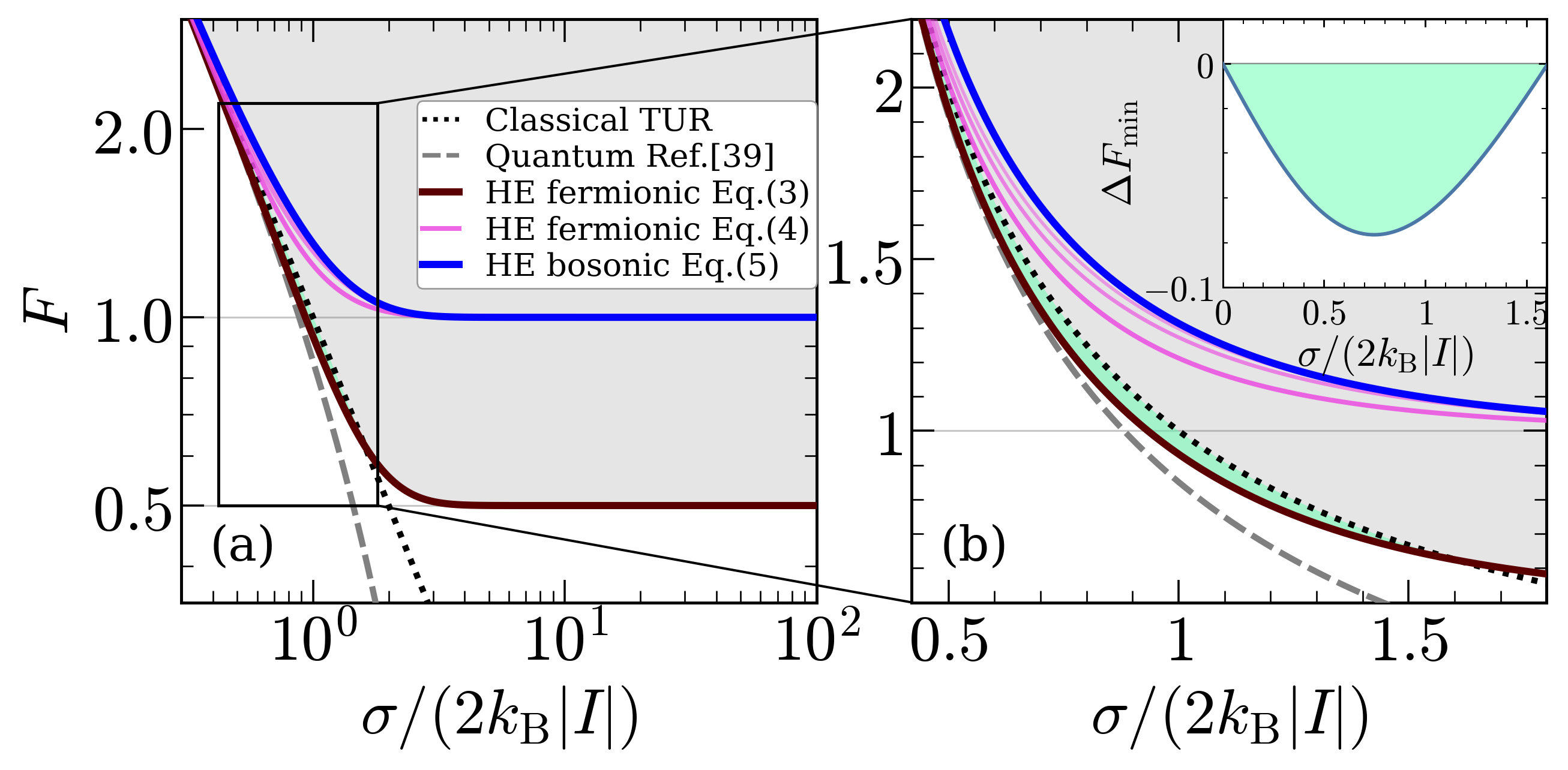}
	\caption{
		 Analytic heat-engine TUR bounds. The brown curve is the fermionic bound of \Cref{eq:HEbound_F}, which saturates at $F\to 1/2$. The pink curves show the
         %$\tau$-dependent fermionic HE bound
         result of \Cref{eq:HEbound_tau}
         for $\tau=3,4,5$.
         %, converging to the bosonic bound as $\tau\to\infty$.
         The blue curve is the bosonic HE bound \Cref{eq:bose_coth}, which saturates at $F\to 1$. The gray shaded area is the region where a fermionic heat engine can operate, while the green shaded area is the region where there is a quantum advantage (see also the inset where we plot the difference $\Delta F_\mathrm{min}$ between the bound~\eqref{eq:HEbound_F} and the classical-TUR bound). The bosonic HE bound is always more restrictive than the classical TUR, offering no quantum advantage.}
	\label{fig2}
\end{figure}
%%%%%%%%%%%%%%%%%%%%%%%%%%%%%%%%%%%%%%%%
For \emph{steady-state} coherent heat engines, a relevant
benchmark is the TUR of Ref.~\cite{brandner2025thermodynamic}, which translates into an efficiency bound that is always less restrictive than the classical one. However, this quantum TUR belongs to the class of results that are valid for a generic thermodynamic process. This means that both the TUR and the associated efficiency constraint can be loose when applied to a heat engine. Thus, the question whether a more informative TUR for steady-state quantum thermal machines can be derived, or equivalently a refined efficiency bound, remains open.

In this work, we precisely answer this question for coherent thermoelectric heat-engines (see Fig.~\ref{fig1}) considering fermionic, bosonic, and classical carriers.
Our key results are as follows. (i) We find new thermodynamic uncertainty relations that can never be violated in coherent transport and impose lower limits on the Fano factor of coherent heat engines. (ii) We provide conditions for which the bounds are saturated, supported by numerical investigation. (iii) We show that our results imply qualitatively different behavior of the efficiency bounds, as compared to the classical TUR, requiring finite fluctuations to operate at nonzero efficiency.

\emph{Fermionic transport---}
For fermions, our first result for a heat-engine efficiency reads
\begin{equation}
\frac{\eta}{\eta_\mathrm{c}} \le \left[\,1 + \frac{k_{\mathrm{B}}T_C}{|\delta\mu|}\, \ln\!\left(\frac{3 + 2\sqrt{F^2+2}}{\,2F-1\,}\right)\right]^{-1}\,.
\label{eq:eff_HE_F}
\end{equation}
Similarly to Eq.~\eqref{eq:efficiency_TUR_cl}, we see that $\eta_\mathrm{c}$ is reached at $F\to\infty$ (either zero power or diverging fluctuations). However, unlike in Eq.~\eqref{eq:efficiency_TUR_cl}, a coherent fermionic heat engine cannot sustain a finite efficiency below the threshold of $F=1/2$.
The bound~\eqref{eq:eff_HE_F} on efficiency descends from an improved, heat-engine-specific TUR which we derive in this work and reads (see End Matter for details)
\begin{equation}
\frac{S_I}{|I|} \ge \frac{1 + \cosh^2\!\big(\sigma/2k_{\mathrm{B}}|I|\big)}{\sinh\!\big(\sigma/k_{\mathrm{B}}|I|\big)}\,.
%:=F_\mathrm{min}^{\mathrm{HE}}\left(\frac{\sigma}{2k_\mathrm{B}|I|}\right).
\label{eq:HEbound_F}
\end{equation}
Precisely due to the heat-engine constraint, Eq.~\eqref{eq:HEbound_F} is strictly tighter than Brandner-Saito's quantum TUR, and reduces to the
classical TUR $S_I \sigma/I^2 \ge 2k_\mathrm{B}$ when $\sigma/2k_\mathrm{B}|I|\ll 1$. At fixed rate of entropy production, the minimum Fano factor attainable by a heat engine always exceeds that of a refrigerator or a dissipator. A key feature of our result is the prediction that classical-TUR
violations available to a coherent fermionic heat engine are far more constrained than those in any other working mode~\cite{supp}.
Indeed, while in general coherent transport permits arbitrary violation of the classical TUR at large biases~\cite{Brandner2018Mar,Ehrlich2021Jul,timpanaro2025quantum,brandner2025thermodynamic}, the lower bound of
Eq.~\eqref{eq:HEbound_F} lies below the classical one only for $0<\sigma/2k_\mathrm{B}|I|<1.606$, with a maximum TUR violation of about $6.78\%$ (relative to the classical TUR) at $\sigma/2k_\mathrm{B}|I|\approx 1.005$
%\redd{, where each electron contributes on average with $2k_B$ to the entropy production rate}.
The right-hand side of Eq.~\eqref{eq:HEbound_F} is a decreasing function of its argument, approaching $1/2$ for $\sigma/2k_\mathrm{B}|I|\gg 1$. Therefore, $F>1/2$ and this behavior is reflected in the form of the upper bound on the efficiency of Eq.~\eqref{eq:eff_HE_F}.
%%%%%%%%%%%%%%%%%%%%%%%%%%%%%%%%%%%%%%%%
\begin{figure}[t]
	\centering
	\includegraphics[width=1\linewidth]{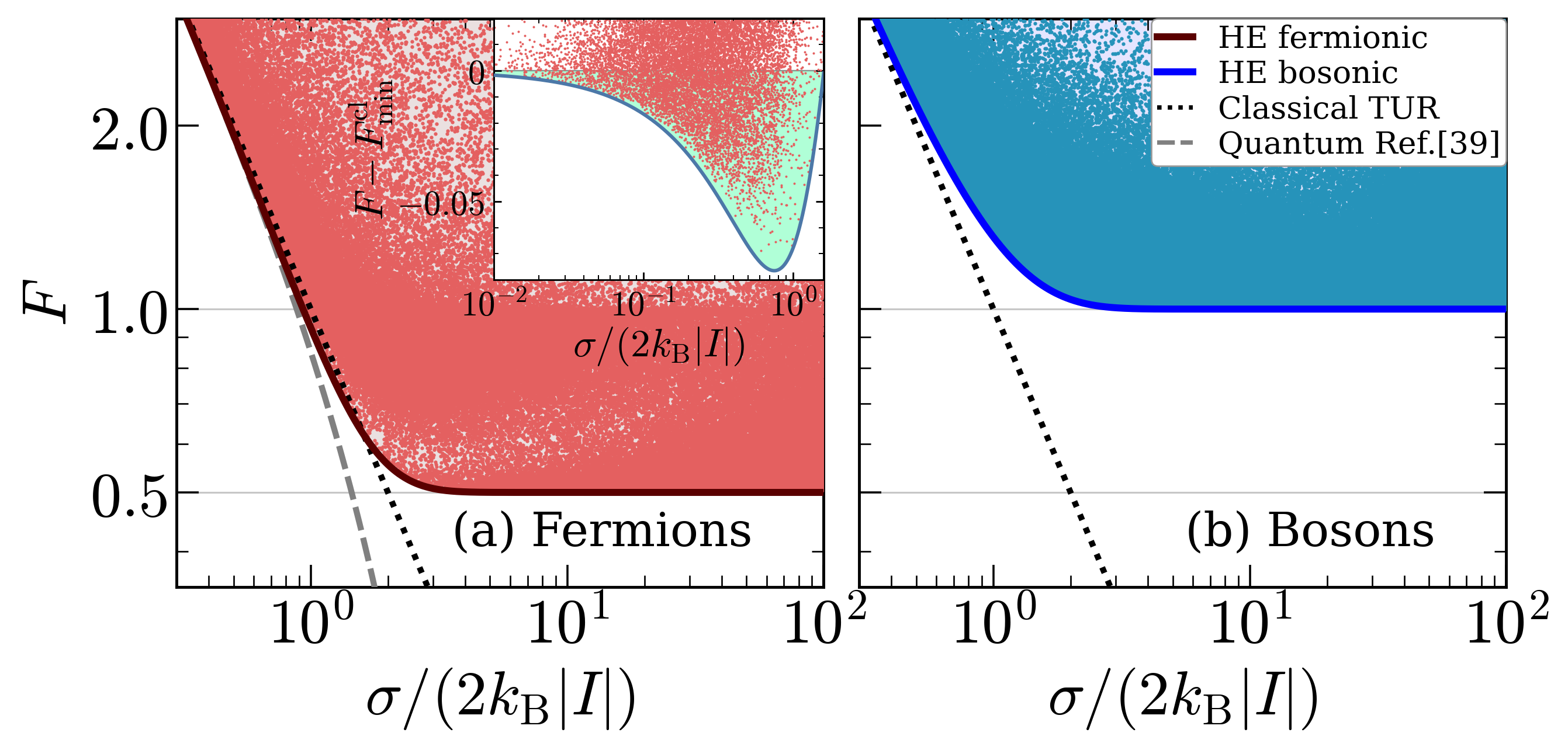}
	\caption{Numerical verification of the TURs derived in this work. (a) Fermions (red). (b) Bosons (blue). The transmission $\mathcal{T}(E)$ is formed by the union of up to three boxcars. Each point is obtained by randomly sampling reservoir temperatures, chemical potentials,  the positions $E_i$ and widths $W_i$ of the boxcars $(i=1,2,3)$, retaining only heat-engine configurations. We use $E_i\in(-500,500), \mu_\alpha\in(-500,500), k_\mathrm{B}T_\alpha\in(10^{-3},500)$, and $W_i\in(10^{-8},500)$. For bosons, $\mu_\alpha\in(-500,-10^{-2})$ and $E_i\in(0,500)$. All parameters are given in an arbitrary energy unit that does not matter because the plotted variables are dimensionless. The gray dashed curve shows for comparison Brandner-Saito fermionic quantum TUR~\cite{brandner2025thermodynamic}, while the black dotted curve is the classical-TUR bound. The inset is a zoom on the region where the classical TUR is violated, showing that we indeed find data points in the green region where the Fano factor goes below the classical-TUR minimum $F_\mathrm{min}^\mathrm{cl}$.
    %In the simulations, the transmission consists of three boxcars of centre $E_i$ and width $W_i$ ($i=1,2,3$), with $T_\alpha\in(10^{-3},500)$, $\mu_\alpha\in(-500,500)$,$E_i\in(-500,500)$, and $W_i\in(10^{-8},300)$, in units of $T_C$. For bosons, $\mu_\alpha\in(-500,-10^{-2})$ and $E_i\in(0,500)$.
    }
	\label{fig3}
\end{figure}
%%%%%%%%%%%%%%%%%%%%%%%%%%%%%%%%%%%%%%%%

Equation~\eqref{eq:HEbound_F} is valid for arbitrary voltage bias, temperature bias, and transmission $\mathcal{T}(E)$, which encodes the probability that a carrier with energy $E$ is transmitted across the scattering region. Nevertheless, when the two reservoirs are close enough in temperature ($T_H\le 2T_C $), it can be further tightened, yielding the temperature-resolved bound
\begin{equation}
	\frac{S_I}{|I|}\ge
	\frac{1+\cosh X\,\cosh(\tau X)}{\sinh X\,\bigl[\cosh X+\cosh(\tau X)\bigr]},
	\quad T_H\le 2T_C 
	\label{eq:HEbound_tau}
\end{equation}
where $X:=\frac{\sigma}{2k_\mathrm{B}|I|}$ and $\tau = \tfrac{T_H + T_C}{T_H - T_C}$. The right-hand side of \Cref{eq:HEbound_tau} is a decreasing function of $X$ that tends to 1 for $X\gg 1$, meaning that the Fano factor of a coherent heat engine with $T_H\le 2T_C$ satisfies $F> 1$. The temperature ratio $T_H/T_C=2$ marks a sharp threshold between two regimes of charge-current precision where the minimum Fano factor jumps from 1 to 1/2. Moreover, the bound of Eq.~\eqref{eq:HEbound_tau} further restricts the region where an advantage over the classical TUR is possible (for instance, $0<X<0.45$ at $T_H=2T_C$).
%always lies above the classical TUR, so no quantum advantage is available in this regime.
Our findings from Eqs.~\eqref{eq:HEbound_F} and~\eqref{eq:HEbound_tau} are illustrated in Fig.~\ref{fig2} by the brown and pink solid lines, respectively. The green shaded area is the region for which a quantum advantage can be achieved.

A natural question is whether the bounds derived here are actually tight, and we show in the following that this is indeed the case.
As demonstrated in Refs.~\cite{timpanaro2025quantum,Danielsson2025Aug}, the current noise $S_I$ is minimized for any number of linear constraints in $\mathcal{T}(E)$ when $\mathcal{T}(E)$ is a collection of boxcar transmissions (with unit height). Such form of $\mathcal{T}(E)$ is then the natural candidate to test our bound, whereas different choices are expected to be far from optimal (in the sense that they do not allow to saturate the bound). In Fig.~\subfigref{fig3}{a} we present a scatter plot obtained by randomly sampling voltage and temperature biases, as well as transmission functions formed as unions of three boxcars (with random positions and widths) when the system acts as a heat engine. The sampled points populate the region above the bound of Eq.~\eqref{eq:HEbound_F} (brown solid line), showing that it is possible to saturate the lower bound even at large dissipation $\sigma\gg 2k_{\mathrm{B}}|I|$,
where kinetic uncertainty relations are usually tighter~\cite{DiTerlizzi2018Dec,Hiura2021May,Vo2022Sep,Macieszczak2024Jul,Prech2025Jan,Prech2025Sep,Palmqvist2025Jul,Palmqvist2025Oct,Blasi2025May}. 
Sufficient conditions for saturation at a given $X$ are that the transmission function selects transport energies where the reservoir occupations are $n_H\approx 1/2$ and $n_C\approx (e^{2X}+1)^{-1}$. At large dissipation, this ensures a large difference in occupation (thus increasing the current and reducing the Fano factor) and minimizes the thermal noise from the cold reservoir. So, for instance, saturation can be reached with a finite-width boxcar centered at $E_0$ just above $\mu_C$, with $\mu_C >\mu_H$ and $T_H\gg T_C$.
When $T_H\le 2T_C$ the condition $n_H\approx 1/2, n_C\approx 0$ is not allowed, and saturation at large $\sigma$ happens for $n_C\ll n_H\ll 1$ (Poissonian electron transport) or $1-n_H\ll 1-n_C\ll 1$ (Poissonian hole transport), explaining why $F\to 1$ in this case.

\begin{figure}[t]
	\centering
	\includegraphics[width=\linewidth]{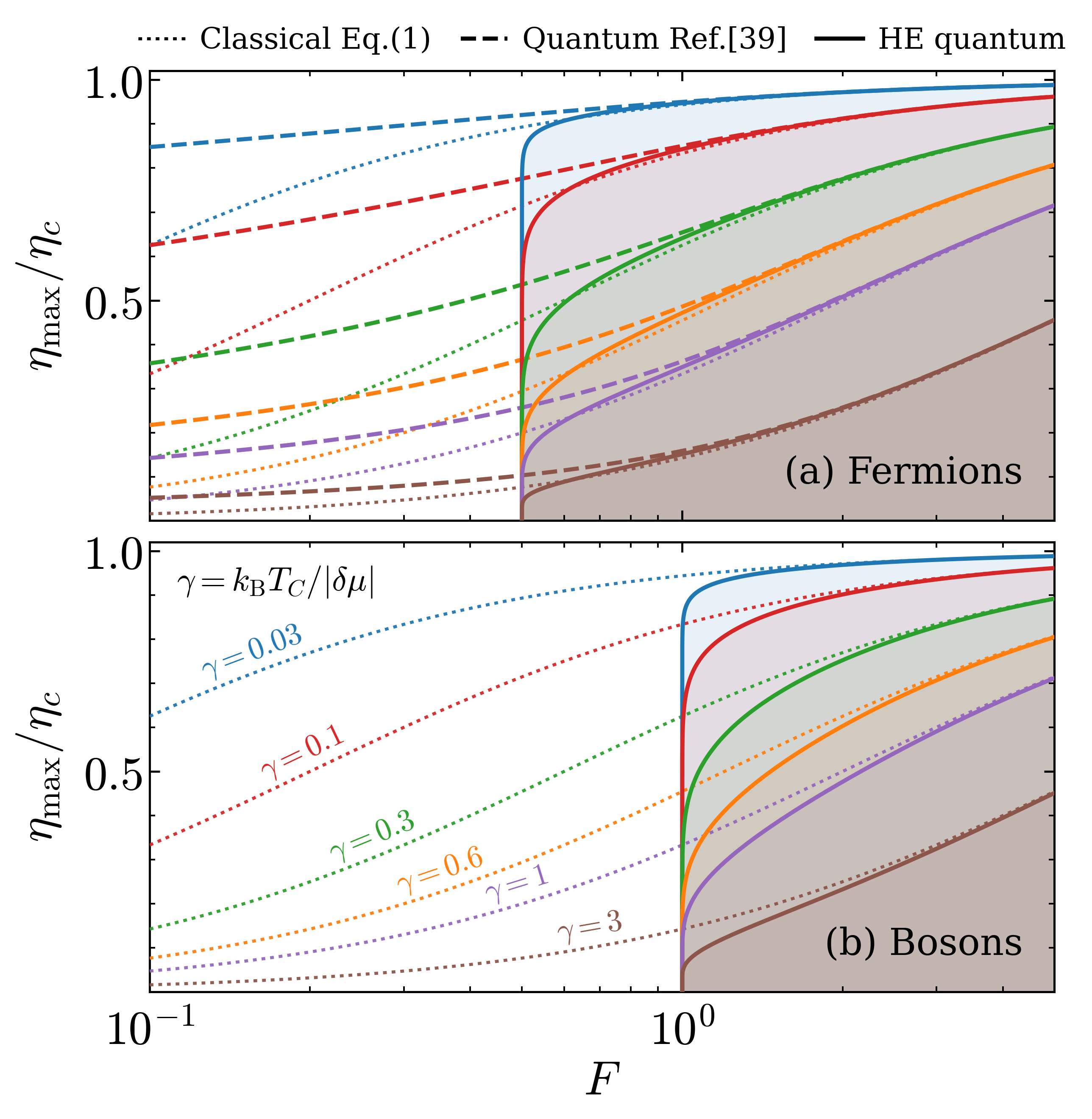}
	\caption{Maximum efficiency $\eta_\mathrm{max}/\eta_\mathrm{c}$ allowed by different bounds as a function of the Fano factor. (a) Fermions. The dotted, dashed, and solid lines correspond to the classical bound~\eqref{eq:efficiency_TUR_cl}, the quantum TUR of Ref.~\cite{brandner2025thermodynamic}, and the heat engine quantum bound~\eqref{eq:eff_HE_F} derived here.
    (b) Bosons. The dotted and solid lines correspond to the classical bound~\eqref{eq:efficiency_TUR_cl} and the bound~\eqref{eq:eff_HE_B} derived here.
    For both panels, different colors indicate different values of the ratio $\gamma=k_\mathrm{B}T_C/|\delta\mu|$.
    }
	\label{fig4}
\end{figure}

The fact that the fermionic TUR of Eq.~\eqref{eq:HEbound_F} can be saturated has the important consequence that the efficiency bound derived from it, Eq.~\eqref{eq:eff_HE_F}, is also tight. This is especially useful because Eq.~\eqref{eq:eff_HE_F} provides an estimate of the maximum efficiency of a coherent, steady-state thermoelectric heat engine in terms of electrical quantities only, without the need to measure heat currents, which is an example of thermodynamic inference~\cite{Gingrich2017Apr,Seifert2019Mar}.
The features of the efficiency bound are inherited from the behavior of the TUR~\eqref{eq:HEbound_F}. First, the maximum allowed efficiency always lies below the previously reported quantum bound of Ref.~\cite{brandner2025thermodynamic} and requires $F > 1/2$.
In terms of output power fluctuation, this condition corresponds to $S_{P}\ge \delta\mu\, P/2 $. This sharp threshold is a novel prediction of Eq.~\eqref{eq:eff_HE_F} that stands in contrast with previous results, like Eq.~\eqref{eq:efficiency_TUR_cl}, where the limit of vanishing maximum efficiency is reached smoothly at vanishing fluctuations $F\to 0$. 
Second, there is only a restricted window $F>0.623$ where the estimate from Eq.~\eqref{eq:eff_HE_F} is slightly above the classical estimate of Eq.~\eqref{eq:efficiency_TUR_cl}, signaling that the quantum advantage is indeed rather limited.
These features are illustrated in Fig.~\subfigref{fig4}{a}, for different values of the ratio $k_\mathrm{B}T_C/|\delta\mu|$. 
In summary, Figs.~\subfigref{fig3}{a} and~\subfigref{fig4}{a} show in a complementary way how constrained the classical-TUR violation is for a fermionic coherent thermoelectric heat engine.
%%%%%%%%%%%%%%%%%%%%%%%%%%%%%%%%
%%%%%%%%%%%%%%%%%%%%%%%%%%%%%%%%
%%%%%%%%%%%%%%%%%%%%%%%%%%%%%%%%
%%%%%%%%%%%%%%%%%%%%%%%%%%%%%%%%

\emph{Bosonic transport---}
In this case, we find the TUR
\begin{equation}
\frac{S_I}{|I|} \ge \coth\!\left(\frac{\sigma}{2k_\mathrm{B}|I|}\right),
\label{eq:bose_coth}
\end{equation}
which is consistent with Ref.~\cite{Palmqvist2025Jul} and always more restrictive than the classical TUR, reducing to it only for $\sigma \ll 2k_\mathrm{B}|I|$. Even though the focus of this work is on heat engines, the above result applies more generally to any two-terminal transport configuration. This result shows that coherent bosonic transport in these conditions can never violate the classical charge-current TUR. In particular, a bosonic heat engine is bounded by $F > 1$ (which can be expected from bunching effects), twice the fermionic one. As for the fermionic bound, the bosonic inequality is tight and can be saturated when $1\gg n_H\gg n_C$. In Fig.~\subfigref{fig2}{b}, one can observe that all the randomly sampled scatter points lie above the bound \Cref{eq:bose_coth}, and saturation can be achieved even at large dissipation. Additionally, this bound corresponds to the $\tau\to\infty$ limit of \Cref{eq:HEbound_tau}. 
The TUR statement~\eqref{eq:bose_coth} can readily be translated into an efficiency bound
%that takes the form
\begin{equation}
\frac{\eta}{\eta_\mathrm{c}} \le \left[\,1 + \frac{k_{\mathrm{B}}T_C}{|\delta\mu|}\, \ln\!\left(\frac{F+1}{F-1}\right)\right]^{-1},
\label{eq:eff_HE_B}
\end{equation}
consistently requiring $F > 1$ and always lying below the classical-TUR constraint~\eqref{eq:efficiency_TUR_cl}. The behavior of the bound~\eqref{eq:eff_HE_B} as a function of the Fano factor is shown in Fig.~\subfigref{fig4}{b} for different values of $k_\mathrm{B}T_C/|\delta\mu|$.

\emph{Classical carriers---} The bosonic results extend to classical ballistic transport, where the carriers obey Maxwell–Boltzmann statistics. Despite this difference in occupation functions, classical and bosonic carriers satisfy the same TUR. Hence, the bounds in~\Cref{eq:bose_coth,eq:eff_HE_B} apply equally to classical ballistic transport.
The fact that bosonic and classical carriers share the same lower bound can be understood by considering the instructive example of a transmission acting as a perfect energy filter, as we now show. We present this case in detail because it allows us to easily derive all the bounds presented in this work and shows how the classical-carrier limit emerges. Moreover, the same perfect-filter results are the building blocks for obtaining the bounds at arbitrary transmission.

\emph{Perfect energy filter---} Consider that $\mathcal{T}(E)$ acts as a perfect filter with unit transmission at $E=E_0$ and width $w_0$ being the smallest energy scale in the problem. In this case, all quantities of interest (currents, noise, entropy production, and Fano factor) can be obtained exactly. To express the results, it is convenient to introduce the variables $x(E)=[a_C(E)-a_H(E)]/2$ and $u(E)=[a_C(E)+a_H(E)]/2$, where $a_\alpha(E)=(E-\mu_\alpha)/k_\mathrm{B}T_\alpha$, with $\alpha\in\{H,C\}$. The quantity $x(E)$ can be interpreted as the energy-resolved entropy production per carrier. Indeed, for the situation of a perfect energy filter that we are considering now, we obtain $X=|x(E_0)|$. By direct calculation from Eqs.~\eqref{eq:current} and~\eqref{eq:noise} (see End Matter), the Fano factor can instead be written as
\begin{equation}
    F=\coth X+2\sum_{n=1}^{\infty}{\left(\zeta\,e^{-u(E_0)}\right)}^n\sinh(nX)\,,
\label{eq:Fano_EF}
\end{equation}
where $\zeta=-1$ for fermions, $\zeta=+1$ for bosons, and $\zeta=0$ for classical particles.
In the latter case, we therefore have $F=\coth X$, so Eq.~\eqref{eq:bose_coth} becomes an equality. Interestingly, this equality agrees with the original result by Barato and Seifert for unicyclic Markov networks in the limiting case of $N=1$ state, corresponding to a biased random walk~\cite{Barato2015Apr,Barato2015May}.
Equation~\eqref{eq:Fano_EF} is very insightful as it shows that the classical-carrier result is obtained from both fermionic and bosonic expressions as $e^{-u(E_0)}\ll 1$, which can be seen as a small parameter for a classical expansion. This is equivalent to say that both occupations $n_\alpha(E_0)$ are small and the reservoirs are made of a very dilute gas, which behaves classically. Indeed, one roughly has $e^{-u(E_0)}\propto \prod_\alpha(\rho_\alpha\lambda_{T_\alpha}^3)^{1/2}$, where $\rho_\alpha$ are the particle densities in the reservoirs and $\lambda_{T_\alpha}$ the thermal wavelength. The standard classical-limit condition $\rho_\alpha\lambda_{T_\alpha}^3\ll 1$ of a dilute gas then implies a small value of $e^{-u(E_0)}$. Compared to classical carriers, bosonic corrections are positive, which explains why the \emph{minimum} Fano factor for bosons coincides with that of classical particles. Fermions produce instead negative corrections, leading to a reduced Fano factor, and consequently a lower minimum. At fixed $X$, the minimum is obtained for $e^{-u(E_0)}=1$ which leads to $F=1/\sinh X$, reproducing the result of Ref.~\cite{brandner2025thermodynamic}. However, this comes from an unconstrained optimization. It turns out that imposing a heat-engine operation (for $I>0$ without loss of generality) requires $u(E_0)>x(E_0)$, so the constrained minimum is achieved at $e^{-u(E_0)}=e^{-X}<1$ (instead of $e^{-u(E_0)}=1$), leading to the bound of Eq.~\eqref{eq:HEbound_F}. Notice that, as soon as the minimization over $u(E_0)$ at fixed $X$ is performed, the parameter $e^{-u(E_0)}$ entering the ``classical expansion'' is lost, which is why the classical-carrier result cannot be obtained from~Eq.~\eqref{eq:HEbound_F}.

Extending the energy-filter results to arbitrary transmission $\mathcal{T}(E)$ for bosons and classical carriers can be done straightforwardly by exploiting the convexity of the energy-filter bound and using Jensen's inequality. For fermions, the proof is more involved.
By exploiting the convexity of the function $y(x(E)):=\frac{1+\cosh^2(x(E))}{\sinh(2x(E))}$ at $x(E)>0$, we are able to show that the energy-resolved Fano factor is lower-bounded at \emph{any energy} by $l_X(x(E))$, where $l_X(z)=p_X+m_X z$ is the tangent line to the function $y$ at point $X$. Integrating this inequality, we can prove the bound of Eq.~\eqref{eq:HEbound_F} for arbitrary transmission, as shown in detail in the End Matter.

\emph{Multiterminal heat engines---} The bound~\eqref{eq:HEbound_F} can be generalized for multiple terminals as follows. Consider $N$ reservoirs, labeled such that $\mu_1\le\mu_2\le\dots\le\mu_N$ and connected by a time-reversal-symmetric coherent conductor, and let $I_\alpha$ be the particle current flowing out of reservoir $\alpha$. We define
%$C_{L_k}=\sum_{\alpha\in L_k}I_\alpha$
$C_k=\sum_{\alpha=1}^{k}I_\alpha$
as the net particle current flowing from the $k$ reservoirs with the lowest chemical potentials
%, $L_k=\{1,\dots,k\}$,
into the $N-k$ remaining ones.
%, $U_k=\{k+1,\dots,N\}$.
The total output power of the system can be written as $P=-\sum_{\alpha=1}^N\mu_\alpha I_\alpha=\sum_{k=1}^{N-1}(\mu_{k+1}-\mu_k)\,C_{k}$,
so that any heat engine ($P>0$) has at least one current $C_{k}>0$ transferring particles against the chemical potential difference defined by the partition into $k$ and $N-k$ reservoirs. For every such current, the associated Fano factor $F_k$ obeys % Eq.~\eqref{eq:HEbound_F} with $I\to C_{L_k}$, $S_I\to S_{L_k}$. 
\begin{equation}
	F_k=\frac{S_{C_k}}{C_{k}} \ge \frac{1 + \cosh^2\!\big(\sigma/2k_{\mathrm{B}}C_{k}\big)}{\sinh\!\big(\sigma/k_{\mathrm{B}}C_{k}\big)}\,,
	\label{eq:HEbound_F_multiterminal}
\end{equation}
and is hence lower-bounded by $1/2$. Ultimately, this relation can be used to obtain a global upper bound on the entropy production as $\sigma\ge\max_{k:\,C_{k}>0}C_{k}\ln[(3+2\sqrt{F_k^2+2})/(2F_k-1)]$. We show in the Supplemental Material~\cite{supp} that the proof follows a similar strategy as the two-terminal one.

\emph{Conclusions and outlook}--- We have established that coherent thermoelectric transport through a  two-terminal conductor with arbitrary transmission function obeys precision limits that depend on the operation mode. For fermionic carriers, imposing heat-engine operation yields a new thermodynamic uncertainty relation: any coherent heat engine satisfies the charge-current TUR of \Cref{eq:HEbound_F}. This new bound confines the violations of the TUR for classical Markov processes to a narrow window of entropy production, limiting them to a few percent, and most strikingly enforces the Fano factor of any fermionic heat engine operating between two reservoirs to be larger than $1/2$. For bosonic and classical carriers the picture changes qualitatively and we obtain a mode-independent charge-current bound~\eqref{eq:bose_coth}, always stricter than the classical TUR. Coherent bosonic transport can therefore never violate it, and is constrained by a Fano factor that is twice the fermionic minimum, $F > 1$. The same limit governs ballistic transport of classical carriers, with the efficiency bound~\eqref{eq:eff_HE_B} applying equally in both cases.
These predictions are experimentally accessible with existing mesoscopic platforms~\cite{Brantut2013Oct,Friedman2020May,Gehring2021Apr,Cangemi2024Oct}.
An interesting direction for future work would be understanding whether the presence of nonthermal distributions in the reservoirs allows one to relax the strong limitations on the fermionic Fano factor found in this work, along the lines provided in Ref.~\cite{Palmqvist2025Jul} for bosons.

\emph{Acknowledgments}--- We are grateful to Kay Brandner, Mart\'i Perarnau--Llobet, Udo Seifert, Janine Splettstoesser, and Fabio Taddei for valuable comments on the manuscript. We acknowledge support from the European
Union under the Horizon Europe research and innovation programme, Marie Sklodowska-Curie grants agreement No. 101148213, EATTS (N.S.) and No. 101205255, FLUTE (M.A.), and by the European Union ERC - RAVE, 101053159 (R.F.). 

\emph{Data availability}---The data that support the findings of this article are openly available in Ref.~\cite{ZenodoData_TURHE_2026}.
%%%%%%%%%%%%%%%%%%%%%%%%%%%%%%%%%%%%%%%%%%%%%%%%%%%%%%%%%%%%%%%%%%%%%%%%%%%%%%%%%%%%%%%%%%%%%%%%%%%%%%%%%%%%%%%%%%%%%%%%%%%%%%%%%%%%%%%%%%%%

\bibliography{biblio}

\clearpage

\begin{center}
\textbf{End Matter}
\end{center}

\setcounter{equation}{0}
\renewcommand{\theequation}{A\arabic{equation}}
\emph{Appendix: details on the derivation---}
We consider a hot reservoir $(T_H,\mu_H)$ and a
cold one $(T_C,\mu_C)$, $T_H \ge T_C$, connected via a coherent central region with arbitrary transmission $0 \le \mathcal{T}(E) \le 1$. The conductor is modeled by scattering theory, allowing one to include Coulomb interactions up to the mean-field level~\cite{Buttiker1993Jun,Christen1996Sep,Whitney2013Aug}. The carriers are either fermions ($\zeta = -1$),
bosons ($\zeta = +1$), or classical particles ($\zeta = 0$), entering the reservoir occupations
$n_\alpha(E) = \{\exp[(E - \mu_\alpha)/T_\alpha] - \zeta\}^{-1}$, with
$\alpha \in \{H,C\}$ ($e = k_\mathrm{B} = 1$ in the following).
We denote the charge and heat currents in a compact form with the variable $\Xi\in\{I,Q_\alpha\}$, where $I=I_H=-I_C$ is the charge current and $Q_\alpha$ the heat currents. They are given by the Landauer--B\"uttiker formulas~\cite{Blanter2000Sep}
\begin{align}
	\Xi
    &=\int \frac{dE}{h}
    \lambda_\Xi(E)
    \mathcal{T}(E)\,\big[n_H(E) - n_C(E)\big],
	\label{eq:current}
\end{align}
where $h$ is Planck's constant, $\lambda_I(E)=1$, and $\lambda_{Q_\alpha}(E)=\nu_\alpha(E-\mu_\alpha)$, with $\nu_{H/C}=\pm 1$, so that $Q_\alpha > 0$ denotes heat flowing out of
reservoir $\alpha$ into the sample. The thermodynamic description in terms of average quantities is completed by introducing the total rate of entropy production $\sigma=-\sum_\alpha Q_\alpha/T_\alpha$ and the output power (work) $P=Q_H+Q_C=\delta\mu\, I$, with $\delta\mu=\mu_C-\mu_H$.

It is convenient to rewrite all the above quantities by introducing the variables $x(E)=[a_C(E)-a_H(E)]/2$ and $u(E)=[a_C(E)+a_H(E)]/2$, where $a_\alpha(E)=(E-\mu_\alpha)/T_\alpha$. With this, we have
\begin{subequations}
\begin{align}
    &\sigma=2\int\frac{dE}{h}\mathcal{T}(E)x(E)g_I(x(E),u(E)),\\
    &g_I(z_1,z_2):=\frac{\sinh z_1}{\cosh z_2+\cosh z_1}\,,
\end{align}
\label{eq:entropy_production}
\end{subequations}
showing that $x(E)$ can be interpreted as the energy-resolved entropy production per carrier. Moreover, the charge current becomes
\begin{align}
    &I=\int\frac{dE}{h}\mathcal{T}(E)g_I(x(E),u(E))\,,
    \label{eq:current_xu}
\end{align}
Fluctuations are captured by the zero-frequency noise,
given by
\begin{subequations}
\begin{align}
	S_\Xi
    &=\int \frac{dE}{h}
    \lambda^2_\Xi(E)
    [\mathcal{S}^\mathrm{th}(E)+\mathcal{S}^\mathrm{sh}(E)]\\
    &=\int \frac{dE}{h}
    \lambda^2_\Xi(E)
    [\mathcal{S}^\mathrm{cl}(E)+\mathcal{S}^\mathrm{qu}(E)]\,.
\end{align}
\label{eq:noise}
\end{subequations}
The integrands define two equivalent decompositions for the noise into thermal and shot or classical and quantum contributions $S_\Xi=S_\Xi^\mathrm{th}+S_\Xi^\mathrm{sh}=S_\Xi^\mathrm{cl}+S_\Xi^\mathrm{qu}$~\cite{Blanter2000Sep}:
\begin{subequations}
\begin{align}
    \mathcal{S}^\mathrm{th}(E)&=\mathcal{T}(E)\sum_\alpha n_\alpha(E)[1+\zeta n_{\alpha}(E)]\,,\\
    \mathcal{S}^\mathrm{cl}(E)&=\mathcal{T}(E)\sum_\alpha n_\alpha(E)[1+\zeta n_{\bar{\alpha}}(E)]\,,
\end{align}
\end{subequations}
with $\bar{\alpha}=C/H$ when $\alpha=H/C$. The remaining components are $\mathcal{S}^\mathrm{qu}=\zeta\mathcal{T}^2(n_H-n_C)^2$ and $\mathcal{S}^\mathrm{sh}=-\zeta\mathcal{T}(1-\mathcal{T})(n_H-n_C)^2$, showing that the quantum component is negative (positive) for fermions (bosons), while the opposite holds for the shot component.
In terms of the $x(E),u(E)$ variables, we have
\begin{subequations}
\begin{align}
    &S_\Xi^\mathrm{th}=\int\frac{dE}{h}\lambda^2_\Xi(E)\mathcal{T}(E)g_S^{\rm th}(x(E),u(E))\,,\\
    &g_S^{\rm th}(z_1,z_2):=\frac{\cosh z_1\cosh z_2-\zeta}{(\cosh z_2-\zeta\cosh z_1)^2},
\end{align}
\label{eq:noiseTH_xu}
\end{subequations}
and
\begin{subequations}
\begin{align}
    &S_\Xi^\mathrm{cl}=\int\frac{dE}{h}\lambda^2_\Xi(E)\mathcal{T}(E)g_S^{\rm cl}(x(E),u(E))\,,\\
    &g_S^{\rm cl}(z_1,z_2):=\frac{\cosh z_1}{\cosh z_2-\zeta\cosh z_1}\,.
\end{align}
\label{eq:noiseCL_xu}
\end{subequations}

Let us start with fermions and fix $\zeta=-1$. First of all, we notice that the output power $P=Q_H+Q_C$ can be written in full generality as $P=(T_H-T_C)u^* I$, where $u^*=(\mu_C-\mu_H)/(T_H-T_C)$ does \emph{not} depend on energy. Moreover, the variables $u(E)$ and $x(E)$ are related by
\begin{equation}
	u(E) = u^{*} + \tau\,x(E),
	\qquad
	\tau = \frac{T_H + T_C}{T_H - T_C} \ge 1.
		\label{eq:uxidentity}
\end{equation}
Consider now a perfect energy filter (EF), characterized by a unit transmission centered at a single energy $E_0$. Then, according to~\eqref{eq:entropy_production} and ~\eqref{eq:current_xu}, $X=\sigma/2|I|=|x(E_0)|$. Since the shot noise vanishes, the Fano factor is $F=S^{\rm th}_I/|I|=y(|x(E_0)|,u(E_0))$ %given in~\eqref{eq:Fano_EF}, which we here express as $F=y(|x(E_0)|,u(E_0))$, 
with the function $y(z_1,z_2):= g_S^{\rm th}(z_1,z_2)/g_I(z_1,z_2)$, see Eqs.~\eqref{eq:current_xu} and ~\eqref{eq:noiseTH_xu}. Imposing that the system operates in a useful working mode (i.e., a heat engine or a refrigerator), leads to the condition $|u(E_0)|\ge |x(E_0)|$ (considering for the heat-engine the least restrictive condition achieved when $\tau=1$).
Across this region, and for fixed $X$, the function $y(X,u(E_0))$
ranges between two limiting curves. The lower envelope, $y_\mathrm{EF}^\mathrm{min}(X)$, is attained at the sector boundary $|u(E_0)| = |x(E_0)|$, where $a_H(E_0) \to 0$ for a heat engine (or $a_C(E_0) \to 0$ for a refrigerator) and one finds the right-hand side of Eq.~\eqref{eq:HEbound_F}
\begin{equation}
	y_{\mathrm{EF}}^{\min}(X) = \frac{1 + \cosh^2 X}{\sinh(2X)}\,.
		\label{eq:EFmin}
\end{equation}
The upper envelope is instead approached as the level moves deep into a sector,
$|u(E_0)| \to \infty$, where $y \to \coth X$. Both curves lie strictly above the quantum
bound of Ref.~\cite{brandner2025thermodynamic}, which is recovered by minimizing the function $y(X,u(E_0))$ over $u(E_0)$ \emph{without any constraint}, as mentioned in the main text. Consequently, for a fixed $X$, the region $y_\mathrm{EF}^\mathrm{min}(X)\le F\le \coth X$ determines the allowed Fano factor accessible to useful working modes.
The dissipative EF region (where both $Q_H<0$ and $Q_C<0$) is instead restricted between Brandner-Saito's quantum TUR and $y_{\mathrm{EF}}^{\min}(X)$, so that the EF lower bound for useful modes acts as the upper bound of the dissipative sector.
For a generic transmission, as we show below, the lower envelope $y_{\mathrm{EF}}^{\min}(X)$
survives as a bound only in the heat-engine mode, and the upper bound disappears.

For arbitrary transmission, we recall that $P=(T_H-T_C)u^* I$.
Heat-engine operation $P>0$ therefore implies $u^*I>0$. Consider first $I>0$, so that $u^*>0$. As the energy $E$ varies, so do the quantities $x(E)$ and $u(E)$, which are however related by~\eqref{eq:uxidentity}, so $u(E)$ moves along a line in the $(x,u)$ plane. On the positive side, $x(E)>0$, using $u^*>0$ one has $u(E)>\tau x(E)\ge x(E)$, and the corresponding energy-resolved points that contribute to the integrals~\eqref{eq:current_xu} and~\eqref{eq:noiseTH_xu} lie in the sector where $y(x(E),u(E))\ge y_{\rm EF}^{\min}(x(E))$. Let
\begin{equation}
    l_X(z)=p_X+m_Xz
\end{equation}
be the tangent line to $y_{\rm EF}^{\min}(z)$ at $X=\sigma/(2I)$. Since $y_{\rm EF}^{\min}(z)$ is positive, convex, and decreasing for $z>0$, we have $m_X<0$, $p_X>0$, and $y_{\rm EF}^{\min}(x(E))\ge l_X(x(E))$, and hence $g_S^{\rm th}(x(E),u(E))\ge l_X(x(E))g_I(x(E),u(E))$ on this branch. On the negative side, $x(E)<0$, the same inequality is automatic: $l_X(x(E))>0$, due to the signs of $p_X$ and $m_X$, while $g_I(x(E),u(E))<0$ and $g_S^{\rm th}(x(E),u(E))\ge0$. So, we have a way to bound $g_S^{\rm th}$ at \emph{any energy}. Using this fact, together with the positivity of the shot noise, we have
\begin{align}
    hS_I&\ge hS_I^{\text{th}}=\int dE \mathcal{T}(E)g_S^{\rm th}(x(E),u(E))\nonumber\\
    &\ge \int dE \mathcal{T}(E)l_X(x(E))g_I(x(E),u(E))\nonumber\\
    &=p_X \int dE \mathcal{T}(E)g_I(x(E),u(E))\nonumber\\
    &\quad+m_X \int dE \mathcal{T}(E)x(E)g_I(x(E),u(E))\nonumber\\
    &= p_X hI+m_X h\sigma/2\,.
\end{align}
Dividing by $I>0$ gives $F\ge l_X(X)=y_\mathrm{min}^\mathrm{EF}(X)$, i.e., Eq.~\eqref{eq:HEbound_F} of the main text. When $I<0$, we consider $\tilde x=-x$ and $\tilde u=-u$, which transforms the $I<0$ sector of the $(x,u)$ plane into the $I>0$ sector of the $(\tilde x,\tilde u)$ plane. Therefore, we can exploit the previous results for the variables $\tilde x$ and $\tilde u$, leading to the inequality $g_S^{\mathrm{th}}(x,u)\ge -l_X(-x)g_I(x,u)$. Repeating the same reasoning as before, we get once again Eq.~\eqref{eq:HEbound_F}, thus completing the proof. 
When $T_H\le 2T_C$, the function
$y(x(E),u(E))$ along the heat-engine boundary $u(E)=\tau x(E)$ is convex for $x(E)>0$, and applying the same tangent-line argument as done above yields the sharper, temperature-resolved bound of Eq.~\eqref{eq:HEbound_tau}.  For $T_H>2T_C$ ($1<\tau<3$) the function $y(x(E),\tau x(E))$ is not convex, and the bound reverts to the $\tau$-independent result of Eq.~\eqref{eq:HEbound_F}.

The same proof does not extend to a refrigerator mode. Consider for instance $I>0$: then $P<0$ implies $u^*<0$. As a result, the line $u(E)=u^*+\tau x(E)$ on the branch $x(E)>0$ \emph{does not} always lie in the region $|u(E)|>|x(E)|$. Thus, refrigerator operation does not impose the pointwise bound $y(x(E),u(E))\ge y_{\rm EF}^{\min}(x(E))$ required by the tangent-line construction. One is instead left with the general pointwise constraint $y(x(E),u(E))\ge 1/\sinh(x(E))$, which gives Brandnder-Saito's bound.

We finally come to bosonic transport ($\zeta = +1$).
In this case, the convenient splitting of the noise is $S_I=S_I^\mathrm{cl}+S_I^\mathrm{qu}$, because $S_I\ge S_I^\mathrm{cl}$ for bosons, with $S_I^\mathrm{cl}$ given in Eq.~\eqref{eq:noiseCL_xu}. The key observation here is that the function $g_S^{\rm cl}$ in Eq.~\eqref{eq:noiseCL_xu} obeys the identity
\begin{equation}
  \tilde{y}(x(E),u(E))=\frac{g_S^{\rm cl}(x(E),u(E))}{|g_I(x(E),u(E))|}=\coth(|x(E)|)
\label{eq:gs_coth_gi}
\end{equation}
for every energy $E$. Notice that the right-hand side of this equality only depends on $x(E)$ and therefore carries no dependence on the working mode. Since $\coth z$ is convex for $z>0$ it is then possible to bound the Fano factor by using Jensen's inequality and that $\coth$ is a decreasing function, leading to
\Cref{eq:bose_coth}:
\begin{align}
    F=&\frac{S_I}{|I|}\ge\frac{S_I^\mathrm{cl}}{|I|}=\frac{\int dE\mathcal{T}(E)g_I(x(E),u(E))\coth(x(E))}{\left|\int dE\mathcal{T}(E)g_I(x(E),u(E))\right|}\nonumber\\
    &\ge\frac{\int dE\mathcal{T}(E)|g_I(x(E),u(E))|\coth(|x(E)|)}{\int dE\mathcal{T}(E)|g_I(x(E),u(E))|}\nonumber\\
    &\ge\coth\left(\frac{\int dE\mathcal{T}(E)|g_I(x(E),u(E))||x(E)|}{\int dE\mathcal{T}(E)|g_I(x(E),u(E))|}\right)\nonumber\\
    &\ge\coth X\,.
\end{align}
For classical carriers, we use the same argument, because Maxwell-Boltzmann occupations also lead to the identity~\eqref{eq:gs_coth_gi} and $S_I=S_I^\mathrm{cl}=S_I^\mathrm{th}$ when $\zeta=0$.

Further details on the bounds, the efficiency reformulations, a comparison with the Fano factor at maximum power, and the multiterminal generalization are provided in the Supplemental Material \cite{supp}.

\end{document}

\setcounter{section}{0}
\setcounter{equation}{0}
\setcounter{figure}{0}
\setcounter{table}{0}

\setcounter{page}{1}

\renewcommand{\thesection}{S\arabic{section}}
\renewcommand{\thetable}{S\arabic{table}}
\renewcommand{\theequation}{S\arabic{equation}}
\renewcommand{\thefigure}{S\arabic{figure}}
\renewcommand{\bibnumfmt}[1]{[S#1]}
\renewcommand{\citenumfont}[1]{S#1}
\DeclareRobustCommand{\nsrev}[1]{{\color{blue}#1}}

\onecolumngrid

\title{Supplemental Material for: Thermodynamic bound on the Fano factor of a coherent thermoelectric heat engine}

\author{Nahual Sobrino}
\affiliation{The Abdus Salam International Centre for Theoretical Physics, Strada Costiera 11, 34151 Trieste, Italy}

\author{Rosario Fazio}
\affiliation{The Abdus Salam International Centre for Theoretical Physics, Strada Costiera 11, 34151 Trieste, Italy}
\affiliation{Dipartimento di Fisica, Universit\`a di Napoli ``Federico II”, Monte S. Angelo, I-80126 Napoli, Italy}
\author{Matteo Acciai}
\affiliation{The Abdus Salam International Centre for Theoretical Physics, Strada Costiera 11, 34151 Trieste, Italy}

\maketitle

\onecolumngrid

\section{Proof of the TUR of a coherent thermoelectric fermionic heat engine}
\label{appA}
In this section, we show more details on the derivation of the improved TUR for fermionic transport, first for an ideal energy filter, and then for an arbitrary transmission function operating as a heat engine. The bounds we derive are inequalities relating the variables
\begin{align}
    X&=\frac{\sigma}{2|I|},\quad    F=\frac{S_I}{|I|}.
\end{align}
In the following, we will make extensive use of the energy-resolved variables
\begin{align}
    x(E)&=\frac{a_C(E)-a_H(E)}{2}\,,\qquad u(E)=\frac{a_C(E)+a_H(E)}{2}\,,
\end{align}
where $a_\alpha(E)=(E-\mu_\alpha)/T_\alpha$ $(k_\mathrm{B}=1)$ enters the fermionic reservoir occupations $n_\alpha(E)=[\text{exp}(a_\alpha(E))+1]^{-1}$, as well as the functions
\begin{subequations}
\begin{align}
    g_I(z_1,z_2)&=\frac{\sinh z_1}{\cosh z_1+\cosh z_2}\label{eq:gI_app},\\
    g_S^{\rm th}(z_1,z_2)&=\frac{1+\cosh z_1\cosh z_2}{(\cosh z_1+\cosh z_2)^2},\\
    y(z_1,z_2)&=\frac{g_S^{\rm th}(z_1,z_2)}{g_I(z_1,z_2)}\,.
\end{align}
\label{eq:gI_gS_y}
\end{subequations}
In terms of these objects, one can rewrite
\begin{align}
    n_H(E)-n_C(E)&=g_I(x(E),u(E))\,,\\
    \sum_\alpha n_\alpha(E)[1-n_\alpha(E)]&=g_S^{\rm th}(x(E),u(E))\,.
\end{align}

Then, the charge and heat currents, entropy production and noise read
\begin{subequations}
\begin{align}
    I &= \int \frac{dE}{h} \mathcal{T}(E)g_I(x(E),u(E)),\\
     Q_\alpha &= \nu_\alpha \int \frac{dE}{h} T_\alpha a_\alpha(E)\mathcal{T}(E)g_I(x(E),u(E)),\\
    \sigma &= 2\int \frac{dE}{h} \mathcal{T}(E)x(E)g_I(x(E),u(E)),\label{eq:sigma_integral}\\
    S_I&=\int \frac{dE}{h}\mathcal{T}(E)g_S^{\rm th}(x(E),u(E))+S_I^\mathrm{sh}\,.
\end{align}
\label{eq:I_sigma_S}
\end{subequations}
 where $\nu_{H/C}=\pm1$, and $S_I^\mathrm{sh}=\int \frac{dE}{h}\mathcal{T}(E)(1-\mathcal{T})g^2_I(x(E),u(E))\ge0$ is the shot noise component.

According to \Cref{eq:gI_app} $x$ and $g_I$ have the same sign, and therefore from \Cref{eq:sigma_integral} follows that $\sigma\ge 0$.
The working modes of the system are determined by the signs of $Q_\alpha$ and $P=Q_C+Q_H$. In particular, the output power reads
\begin{align}
    P&=\int\frac{dE}{h}\mathcal{T}(E)[T_H a_H(E)-T_C a_C(E)]g_I(x(E),u(E))=(T_H-T_C)\int\frac{dE}{h}\mathcal{T}(E)[u(E)-\tau x(E)]g_I(x(E),u(E))
\end{align}
where $\tau=(T_H+T_C)/(T_H-T_C)\ge 1$. Crucially the quantity
\begin{equation}
    u^*\equiv u(E)-\tau x(E)=\frac{\mu_C-\mu_H}{T_H-T_C}
\label{eq:ustar}
\end{equation}
is \emph{independent} of $E$. Therefore, we can pull it out of the integral and write
\begin{equation}
    P=(T_H-T_C)u^*I\,.
\label{eq:pout_ustar}
\end{equation}
This expression shows that $u^*$ plays a crucial role in determining the sign of the output power. It is worth noting that $u^*=u(E^*)$, where
\begin{equation}
    E^*=\frac{\mu_C T_H-\mu_H T_C}{T_H-T_C}
\end{equation}
is the reversible energy at which $n_H(E^*)=n_C(E^*)$ and $x(E^*)=0$, so no entropy is produced when transport occurs at this energy.

\subsection{Ideal energy filter}
\label{app:single_energy_bounds}
We start the analysis from the case of an ideal energy filter, modeled as a unit transmission at $E=E_0$ with width $w_0\to 0$ being the smallest energy scale in the problem. The general expressions~\eqref{eq:I_sigma_S} become
\begin{subequations}
    \begin{align}
        I&\!=\frac{w_0}{h}g_I(x(E_0),u(E_0))\,,\\
        Q_\alpha&\!=\frac{w_0}{h}\nu_\alpha T_\alpha a_\alpha(E_0)g_I(x(E_0),u(E_0))\,,\\
        \sigma&=\!\frac{2w_0}{h}x(E_0)g_I(x(E_0),u(E_0))=2x(E_0)I,\label{eq:sigma_positivity_appA}\\
        S_I&\!=\frac{w_0}{h}g_S^{\rm th}(x(E_0),u(E_0))\,.
    \end{align}
\end{subequations}
Then, we immediately see that $X=|x(E_0)|$, and, using Eq.~\eqref{eq:gI_gS_y}, $F= y(X,u(E_0))$. This expression is even in $u(E_0)$ and, since $X\ge 0$,
\begin{equation*}
	\frac{\partial y(X,u(E_0))}{\partial \cosh u(E_0)}
	=
	\frac{\sinh X}{(\cosh u(E_0)+\cosh X)^2}\ge 0\,.
\end{equation*}
Thus, $y(X,u(E_0))$ increases monotonically in $|u(E_0)|$.

The useful working modes are characterized by the signs of the heat currents and of the output power, the latter given in~\eqref{eq:pout_ustar}, see for instance Refs.~\cite{buffoni2019quantum} and \cite{Vidal2026May} for a detailed analysis on the working modes. 
Since $I$ and $x(E_0)$ have the same sign due to Eq.~\eqref{eq:sigma_positivity_appA}, we get
\begin{equation}
	P>0
	\iff
    \begin{cases}
        u(E_0)>\tau x(E_0)=\tau X & I>0\\
        u(E_0)<\tau x(E_0)=-\tau X & I<0
    \end{cases}\,.
	\label{eq:HE_condition_single_energy}
\end{equation}
This automatically implies $Q_H>0$ and $Q_C<0$, namely a heat-engine operating mode. Since $\tau\ge 1$, the heat-engine region is a subregion of
the branch $u(E_0)>x(E_0)$ (for $I>0$) or $u(E_0)<x(E_0)$ (for $I<0$), reducing to it when $T_C=0$.

\begin{figure}[t]
    \centering
    \includegraphics[width=0.85\linewidth]{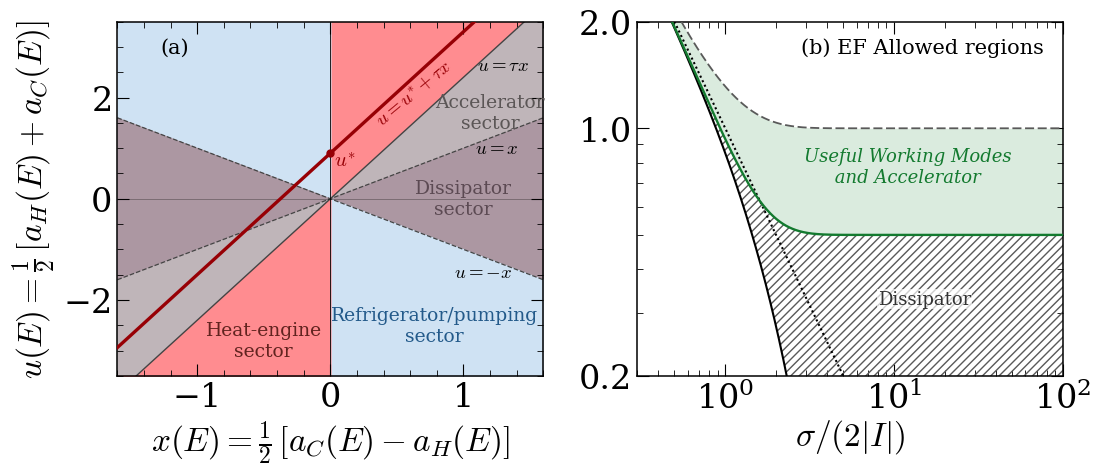}
    \caption{(a) Map of the regions in the $(x(E),u(E))$ plane corresponding to different working modes for the \emph{single-energy} problem: heat engine, refrigerator, accelerator, and dissipator. The boundary between Heat engine and the Accelerator regions is $u=\tau x$, so the Accelerator region disappears for $\tau=1$. The solid dark red line represent a particular evolution of the $u(E)$ variable---which follows a line as the energy is varied---for fixed $\tau=2.4$, and $u^{*}=0.9$.(b) Allowed regions for a perfect energy filter: The Useful working modes (Heat engine and Refrigerator) together with Accelerator mode, lie in the green shaded region given by \Cref{eq:bound_EF_app}, while the dissipator region corresponds to \cref{eq:bound_EF_dissip_app}.}
    \label{fig:S1}
\end{figure}

\begin{figure}[t]
    \centering
    \includegraphics[width=0.85\linewidth]{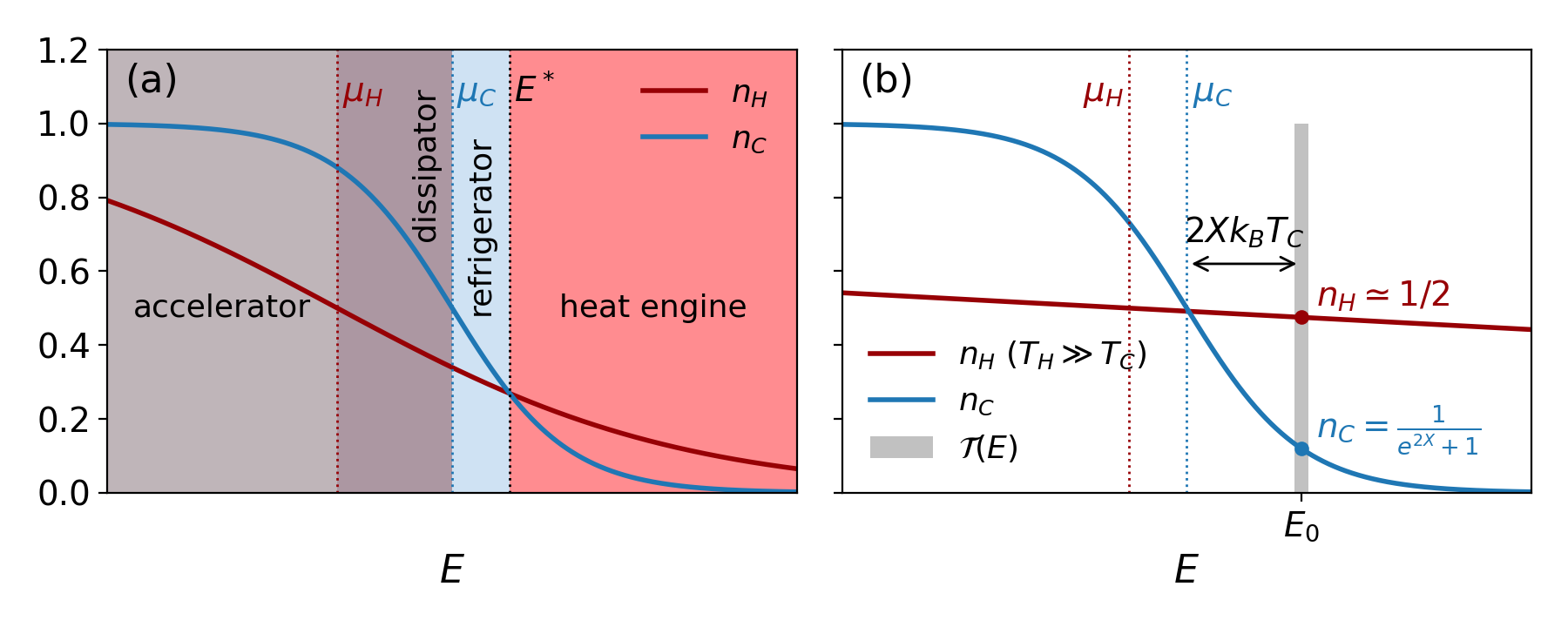}
    \caption{(a) Operation modes of a single energy filter for $\mu_H<\mu_C$. (b) Saturating configuration of \cref{eq:HE_TUR_app}. A narrow boxcar trnsmission is centered at $E_0=\mu_C+2Xk_BT_C$ with $\mu_C>\mu_H$ and $T_H\gg T_C$. }
    \label{fig:Snew}
\end{figure}

For any $T_C\neq 0$, the regions $x(E_0)<u(E_0)<\tau x(E_0)$ (for $I>0$) or $\tau x(E_0)<u(E_0)<x(E_0)$ (for $I>0$) correspond instead to $Q_H>0$, $Q_C<0$, but $P<0$ (accelerator), therefore heat flows from the hot to the cold reservoir, while work is consumed to increase this heat flow.
For a refrigerator operation, the condition $Q_C>0$ is equivalent to
\begin{equation}
    Q_C>0\iff
    \begin{cases}
        u(E_0)<-x(E_0)=-X & I>0\\
        u(E_0)>-x(E_0)=X &I<0
    \end{cases}\,.
\end{equation}
This, in turn, implies $Q_H<0$ and $P<0$, consistently with refrigerator mode. Complementary to all the regions identified above is the sector defined by $|u(E_0)|<|x(E_0)|$, where $Q_C$, $Q_H$, and $P$ are all negative, and the system acts as a dissipator. The situation is summarized in Fig.~\ref{fig:S1}(a) and Fig.~\ref{fig:Snew}(a).

From this analysis, it follows that the \emph{useful} working modes, namely heat engine and refrigerator, together with the accelerator mode, lie in the cone $|u(E_0)|>|x(E_0)|$ in the $(x,y)$ plane (considering the worst-case scenario $\tau=1$ which gives the less restrictive region for the heat engine. So, we conclude that a useful working mode implies $|u(E_0)|>X$. In this region, the function $y(X,u(E_0))$ varies between a lower and a upper envelope, obtained at the two extremes $|u(E_0)|=X$ and $|u(E_0)|\to\infty$. In the first case, corresponding to either $a_H(E_0)\to 0^\pm$ of the $Q_H>0$ branch or $a_C(E_0)\to 0^\mp$ of the refrigerator branch (where the upper and lower signs in the limits refer to $I>0$ and $I<0$, respectively), we get
\begin{equation}
    \lim_{|u(E_0)|\to X^\pm} y(X,u(E_0))=\frac{1+\cosh^2 X}{\sinh(2X)}\,.
    \label{eq:limX_y_EF}
\end{equation}
In the opposite limit, we have
\begin{equation}
    \lim_{|u(E_0)|\to \infty} y(X,u(E_0))=\coth X\,.
\end{equation}
Gathering the results, we obtain the bound
\begin{equation}
    \frac{1+\cosh^2 X}{\sinh(2X)}\le Y \le \coth X
\label{eq:bound_EF_app}
\end{equation}
which is valid for $|u(E_0)|\ge X$, i.e., for a heat engine, a refrigerator or an accelerator. For the stricter, $\tau$-dependent heat engine region, the lower bound translates instead to
\begin{equation}
    F\ge y(X,\tau X)\,,
\end{equation}
with the functional form of $y(u,X)$ given by~\eqref{eq:gI_gS_y}. The upper bound $F\leq\coth X$
remains unchanged. The remaining case is the dissipator region where $|u(E_0)|<X$ . In this situation 
 the lower bound reduces to Brandner-Saito result, while the upper bound corresponds to \cref{eq:limX_y_EF}, and therefore 
\begin{equation}
    \frac{1}{\sinh X}\le F \le \frac{1+\cosh^2X}{\sinh(2X)}.
    \label{eq:bound_EF_dissip_app}
\end{equation}
It is important to emphasize that, for the single-energy problem, the lower boundary in~\eqref{eq:bound_EF_app} is obtained in the limit $n_H\to 1/2$ for a heat engine and $n_C\to 1/2$ for a refrigerator. Regarding the heat engine, this can either be achieved when $E_0\to\mu_H$, in which case $P\to 0$ as well, or for a temperature $T_H$ that is much larger than the all other energy scales relevant for the transport.

The allowed regions for a fermionic perfect energy filter are shown in \cref{{fig:S1}}(b).

%%%%%%%%%%%%%%%%%%%%%%%%%%%%%%%%
%%%%%%%%%%%%%%%%%%%%%%%%%%%%%%%%

\subsection{General transmission}
\label{appB}
In Sec~\ref{app:single_energy_bounds} we showed that the single-energy heat engine has a lower envelope characterized by a function of the form
\begin{align}
    y_{\text{EF}}^{\text{min}}(z)=\frac{1+\cosh^2z}{\sinh(2z)}.
\label{eq:yEF_app}
\end{align}
Here, we show that this bound survives in the heat engine sector for an arbitrary transmission.
In the following, we first consider the case $I>0$.
The function~\eqref{eq:yEF_app} is decreasing and convex for $z>0$. Therefore, its tangent line at $X$
\begin{align}
    l_X(z)=p(X)+m(X)z,
\label{eq:l_tangent}
\end{align}
with 
\begin{subequations}
\begin{align}
    m(X)&=y_{\mathrm{EF}}^{\mathrm{min}\prime}(X)<0,\\
    p(X)&=y_{\mathrm{EF}}^{\mathrm{min}}(X)-m(X)X>0,
\end{align}
\label{eq:m_p_tangent}
\end{subequations}
satisfies 
\begin{align}
    y_{\text{EF}}^{\text{min}}(z)&\ge l_X(z) \quad \forall z>0\,.
\end{align}
The key step of the proof is to observe that the inequality
\begin{align}
    g_S^{\rm th}(x(E),u(E))\ge l_X(x(E))g_I(x(E),u(E)),
    \label{eq:ineq_gS_lXgI}
\end{align}
holds for all energies when the system acts as a heat engine. Indeed, once \cref{eq:ineq_gS_lXgI} is proven, the result follows immediately, because in the noise expression we can bound the integrand $g_S^{\rm th}$ by a linear function in $x(E)$, thus making the entropy production appear in the final expression. Explicitly, \cref{eq:ineq_gS_lXgI} gives
\begin{align}
    h S_I^{\text{th}}&=\int dE \mathcal{T}(E)g_S^{\rm th}(x(E),u(E))\ge \int dE \mathcal{T}(E)l_X(x(E))g_I(x(E),u(E))\nonumber\\
    &=p(X) \int dE \mathcal{T}(E)g_I(x(E),u(E))+m(X) \int dE \mathcal{T}(E)x(E)g_I(x(E),u(E))= p(X)hI+m(X)h\sigma/2.
\end{align}
Making use that for fermions $S_I\ge S_I^{\text{th}}$ and dividing by $I>0$, one obtains
\begin{align}
    \frac{S_I}{I}\ge p(X)+m(X)X=l_X(X)= y_{\text{EF}}^{\text{min}}(X).
    \label{eq:bound_arbit_T}
\end{align}
It remains to prove \cref{eq:ineq_gS_lXgI}.
For this, we exploit Eqs.~\eqref{eq:ustar} and~\eqref{eq:pout_ustar}. They tell us that, for $I>0$, heat-engine mode with $P>0$ implies $u^*>0$ and therefore $u(E)>\tau x(E)\ge x(E)$ at any energy $E$. We now distinguish between the two possible cases.\\
\emph{(i)} If $g_I(x(E),u(E))>0$, then from \cref{eq:gI_app} $x(E)> 0$, and so $E$ lies in the heat-engine sector of the single-energy problem of Sec.~\ref{app:single_energy_bounds}. Therefore 
\begin{align}
    \frac{g_S^{\rm th}(x(E),u(E))}{g_I(x(E),u(E))}\ge y_{\text{EF}}^{\text{min}}(x(E))\ge l_X(x(E)).
\end{align}
\emph{(ii)} If $g_I(x(E),u(E))<0$, then $x(E)< 0$ and therefore $l_X(x(E))> 0$ by Eqs.~\eqref{eq:l_tangent} and~\eqref{eq:m_p_tangent}, so the RHS of \cref{eq:ineq_gS_lXgI} is negative. Since $g_S^{\rm th}(x(E),u(E))\ge 0$ it automatically follows that \cref{eq:ineq_gS_lXgI} holds. 

When $I<0$ we use that
\begin{equation}
    g_S^{\rm th}(x(E),u(E))\ge-l_X(-x(E))g_I(x(E),u(E))
\label{eq:ineq_gS_gI_lX_complementary}
\end{equation}
for all energies. Defining $\tilde x=-x$ and $\tilde u=-u$, the above inequality becomes exactly \cref{eq:ineq_gS_lXgI} in the variables $\tilde x$ and $\tilde u$. Moreover, the heat-engine sector at $I<0$ in the $(x,y)$ plane becomes the heat-engine sector at $I>0$ in the plane $(\tilde x,\tilde u)$, allowing us to exploit the previous proof and establish \cref{eq:ineq_gS_gI_lX_complementary}. Using the latter, we get 
\begin{align}
    hS_I&\ge -\int dE \mathcal{T}(E)l_X(-x(E))g_I(x(E),u(E))\nonumber\\
    &=-p(X) \int dE \mathcal{T}(E)g_I(x(E),u(E))+m(X) \int dE \mathcal{T}(E)x(E)g_I(x(E),u(E))\nonumber\\
    &= -p(X)hI+m(X)h\frac{\sigma}{2}=p(X)h|I|+m(X)h\frac{\sigma}{2},
\end{align}
from which the final result follows by dividing by $|I|$. We have therefore shown that, for a heat engine, 
\begin{align}
    \frac{S}{|I|}\ge  y_{\text{EF}}^{\text{min}}\left(\frac{\sigma}{2|I|}\right)=\frac{1+\cosh^2(\sigma/2|I|)}{\sinh(\sigma/|I|)},
    \label{eq:HE_TUR_app}
\end{align}

The bound just derived uses the convexity of $y_{\mathrm{EF}}^{\min}(x)=y(x,x)$, which lower-bounds $y(x,u)$ over the entire heat-engine cone $u\ge\tau x\ge x$. A configuration that saturates Eq.~\eqref{eq:HE_TUR_app} is illustrated in Fig,~\ref{fig:Snew}(b), consisting in a narrow boxcar transmission  $\mathcal{T}(E)$ is centered at $E_0=\mu_C+2Xk_\mathrm{B}T_C$, with
$\mu_C>\mu_H$ and $T_H\gg T_C$, so that $n_H(E_0)\simeq 1/2$ and
$n_C(E_0)=1/(e^{2X}+1)$. In this limit $F\to y^{\min}_{\rm EF}(X)$.

A stronger, temperature-dependent bound can be obtained by lower-bounding the integrand directly with $y$ evaluated on the boundary line $u=\tau x$. Since $y(z_1,z_2)$ increases with $\cosh z_2$ (i.e.\ with $|z_2|$), every current-carrying level ($x(E)>0$, $u(E)=u^*+\tau x(E)\ge\tau x(E)$ for a heat engine) obeys
\begin{equation}
	y(x(E),u(E))\ge y(x(E),\tau x(E))\equiv y_\tau(x(E)),
	\label{eq:y_ge_htau}
\end{equation}
with
\begin{equation}
	y_\tau(z)=\frac{1+\cosh z\,\cosh(\tau z)}{\sinh z\,[\cosh z+\cosh(\tau z)]}\,.
	\label{eq:ytau_def}
\end{equation}
This function is decreasing and \emph{convex} on $(0,\infty)$ if and only if $\tau=1$  ($T_C=0$) or if $\tau\ge3$ ($T_H\le 2T_C$). When the latter holds, we can define a tangent line $l_X^\tau(z)=p_\tau(X)+m_\tau(X)z$ with $m_\tau(X)<0$ and $p_\tau(X)>0$, and the steps in Eqs.~\eqref{eq:l_tangent}--\eqref{eq:ineq_gS_lXgI} carry over analogously with the replacement $y_{\mathrm{EF}}^{\min}\to y_\tau$, using \cref{eq:y_ge_htau} in place of the cone bound. This yields
\begin{equation}
	\frac{S_I}{|I|}\ge y_\tau\!\left(\frac{\sigma}{2|I|}\right)\ge 1\qquad(\tau\ge3)\,,
	\label{eq:HEbound_tau_app}
\end{equation}
For $1<\tau<3$ ($T_H>2T_C$) the function $y_\tau$ is not convex, the tangent-line step fails, and the bound reverts to the $\tau$-independent envelope $y_{\mathrm{EF}}^{\min}$ of \cref{eq:HE_TUR_app}, with floor $1/2$ (recovered as $\tau\to1$, where $h_\tau\to y_{\mathrm{EF}}^{\min}$).

The proof does not extend in the same way to the refrigerator mode.  To see this, consider for simplicity the case $I>0$. Since $P=(T_H-T_C)u^*I$, refrigerator operation, $P<0$, requires $u^*<0$. Unlike in the heat-engine case, the line $u(E)=u^*+\tau x(E)$ is not confined on the positive-current branch $x(E)>0$ to the useful single-energy cone where $y(x,u)\ge y_{\rm EF}^{\min}(x)$. Instead, it intersects the boundary $u=-x$ at $x=-u^*/(\tau+1)$ and subsequently passes through the dissipator sector before reaching the boundary $u=x$ at $x=-u^*/(\tau-1)$, for $\tau>1$. Therefore, the global refrigerator conditions do not enforce the pointwise inequality $g_S^{\rm th}(x(E),u(E))/g_I(x(E),u(E)) \ge y_{\rm EF}^{\min}(x(E))$ on all current-carrying energies. The tangent-line argument used for the heat engine is consequently not available for arbitrary transmissions. At most, one can use the mode-independent pointwise bound leading to Brandner-Saito's TUR (incidentally, this is a proof of Brandner-Saito's result for two terminals using a different strategy than in the original paper).

\begin{figure}[t]
	\centering
	\includegraphics[width=\linewidth]{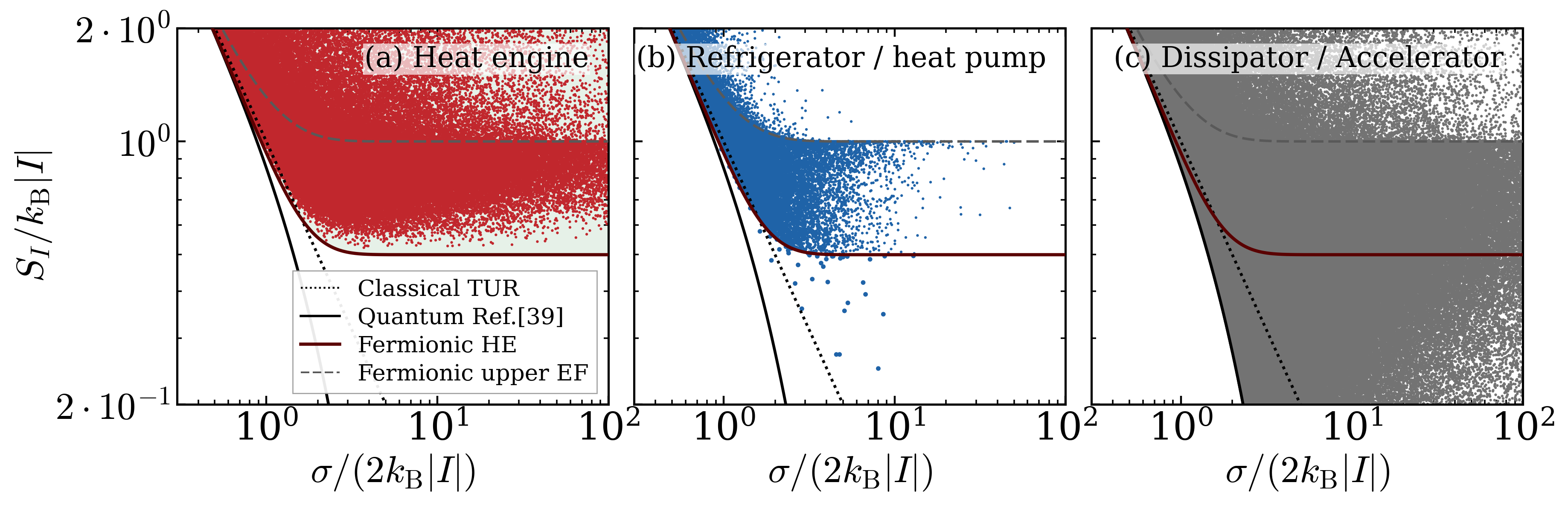}
	\caption{Charge-current TUR ratios for transmission $\mathcal{T}(E)$ formed by the union of up to three boxcars. Each point in the plot is obtained by randomly sampling temperatures, chemical potentials, and the position/widths of the boxcars. (a) Heat-engine mode red); (b) refrigerator/heat-pump (blue); (c) dissipator and accelerator (gray). The black solid curve is the Brandner-Saito bound, the black dotted curve is the classical TUR, and the gray dashed curve is the upper bound for a perfect energy filter. The brown solid curve denotes the heat-engine quantum lower bound of \cref{eq:HE_TUR_app}.  }
	\label{figS2}
\end{figure}

In Fig.~\ref{figS2} we present the fermionic TUR ratios obtained by randomly sampling voltage and temperature biases, as well as transmission functions formed as unions of three boxcars (with random positions and widths). The resulting points, classified according to their working mode, populate the region above \cref{eq:bound_arbit_T} (brown solid line) in panel~(a), while refrigerator~(b) and dissipative/ accelerator~(c) points descend to Brandner-Saito's quantum bound. While this is evident for dissipator configurations, only a few refrigerator points appear to approach the lower quantum bound. This is because doing so requires a transmission made of at least two different boxcars with rather fine-tuned positions, which is rarely found in our wide sampling range.
\subsection{Comparison with the Fano factor at maximum power}
From Ref.~\cite{Whitney2015Mar}, we know that there is a maximum power for a fermionic thermoelectric heat engine, which is
\begin{equation}
    P_\mathrm{max}\propto \frac{1}{h}(T_H-T_C)^2
\end{equation}
where the proportionality constant is obtained by an optimization process as follows. First, the transmission function that allows one to get such a maximum power is a step function $\Theta(E-\epsilon_0)$ with onset at the reversible energy
\begin{equation}
    \epsilon_0=\frac{\mu_C T_H-\mu_H T_C}{T_H-T_C}
\end{equation}
that is the (one and only) energy where the two Fermi functions cross. Then, one can calculate the power and look for the optimal voltage bias $\Delta\mu_{\mathrm{max}}$ that maximizes it. The power turns out to be
\begin{equation}
    P=\frac{1}{h}\Delta\mu\Delta T\ln(1+e^{-\Delta\mu/\Delta T})
\end{equation}
with $\Delta\mu=\mu_C-\mu_H$ and $\Delta T=T_H-T_C>0$. Setting $\partial P/\partial(\Delta\mu)=0$ one finds the following equation in the variable $\delta_0=\Delta\mu/\Delta T$:
\begin{equation}
    \delta_0 e^{-\delta_0}=(1+e^{-\delta_0})\ln(1+e^{-\delta_0})\,.
\end{equation}
This gives the optimal bias as $\Delta\mu_{\mathrm{max}}=\delta_0\Delta T$. The numerical factor is found to be $\delta_0\approx 1.146$.
With these findings, the current associated with the maximal power is
\begin{equation}
    I_\mathrm{max}=\frac{1}{h}\Delta T\ln(1+e^{-\delta_0})
\end{equation}
and
\begin{equation}
    P_{\mathrm{max}}=\frac{1}{h}\delta_0\ln(1+e^{-\delta_0})(T_H-T_C)^2\,.
\label{eq:pmax}
\end{equation}
At this optimal point, the noise can also be evaluated explicitly, yielding
\begin{equation}
    S_{I_\mathrm{max}}=\frac{1}{h}\int_{\epsilon_0}^{+\infty}dE\sum_\alpha f_\alpha(1-f_\alpha)=\frac{T_H+T_C}{h}\frac{1}{1+e^{\delta_0}}\,.
\end{equation}
This leads to a Fano factor of
\begin{equation}
    F|_{P_{\mathrm{max}}}=\frac{T_H+T_C}{T_H-T_C}\frac{1}{(1+e^{\delta_0})\ln(1+e^{-\delta_0})}\equiv B_0 \frac{T_H+T_C}{T_H-T_C}\,.
\end{equation}
The minimum Fano factor is obtained in the limit $T_H\gg T_C$ and is given by $\min_{T_H,T_C} F|_{P_\mathrm{max}}=B_0\approx 0.873$.

\begin{figure}
    \centering
    \includegraphics[width=0.45\linewidth]{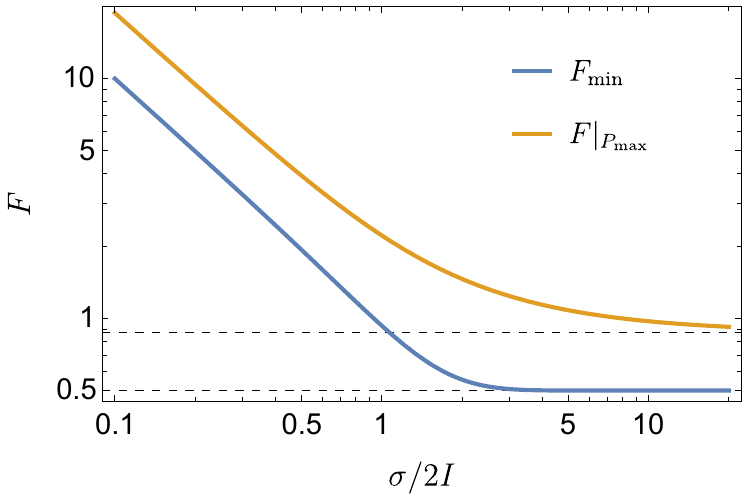}
    \caption{Comparison between the minimum achievable Fano factor (as predicted by our TUR) and the Fano factor at maximum power $F|_{P_\mathrm{max}}$. Notice that while $F_\mathrm{min}$ is a lower bound, when the system operates at maximum power, the Fano factor is fixed and always follows the orange curve.}
    \label{fig:Fano_pmax}
\end{figure}

The entropy production can also be evaluated explicitly. We start from the expression
\begin{equation}
    \sigma=I\left(\frac{\mu_H}{T_H}-\frac{\mu_C}{T_C}\right)+I_E\left(\frac{1}{T_C}-\frac{1}{T_H}\right)\,.
\label{eq:sigma}
\end{equation}
When we consider the transmission that gives the maximum power, the integrals can be evaluated explicitly, yielding
\begin{equation}
    I_E^\mathrm{max}=F_H(\epsilon_0)-F_C(\epsilon_0)\,,
\end{equation}
with
\begin{equation}
    hF_\alpha(\epsilon_0)=\epsilon_0 T_\alpha\ln(1+e^{-\frac{\epsilon_0-\mu_\alpha}{T_\alpha}})-T_\alpha^2\mathrm{Li}_2(-e^{-\frac{\epsilon_0-\mu_\alpha}{T_\alpha}})\,,
\end{equation}
where $\mathrm{Li}_2$ denotes the dilogarithm function. Recalling that $\epsilon_0=\mu_H+\delta_0 T_H=\mu_C+\delta_0T_C$, we get
\begin{equation}
    hF_\alpha(\epsilon_0)=\epsilon_0 T_\alpha\ln(1+e^{-\delta_0})-T_\alpha^2\,\mathrm{Li}_2(-e^{-\delta_0})\,.
\end{equation}
Therefore
\begin{equation}
    h I_E^\mathrm{max}=\epsilon_0(T_H-T_C)\ln(1+e^{-\delta_0})-(T_H^2-T_C^2)\mathrm{Li}_2(-e^{-\delta_0})
\end{equation}
We now want to express the Fano factor as a function of the TUR variable $X=\sigma/2I$. Starting from Eq.~\eqref{eq:sigma} we have
\begin{align}
    X_\mathrm{m}&:=\left.\frac{\sigma}{2I}\right|_{P_\mathrm{max}}=\frac{\mu_H T_C-\mu_C T_H}{2T_H T_C}+\frac{I_E}{2I}\frac{T_H-T_C}{T_H T_C}=-\frac{\epsilon_0(T_H-T_C)}{2T_H T_C}+\frac{hI_E}{2T_HT_C\ln(1+e^{-b_0})}\nonumber\\
    &=-\frac{T_H^2-T_C^2}{2T_HT_C}\frac{\mathrm{Li}_2(-e^{-\delta_0})}{\ln(1+e^{-\delta_0})}\equiv A_0 \frac{T_H^2-T_C^2}{T_HT_C}\,.
\end{align}
The numerical prefactor is $A_0\approx 0.536$.
Setting $\tilde T_{H}\equiv T_H/T_C$, we can invert this relation to find
\begin{equation}
    \tilde T_H=\frac{X_\mathrm{m}+\sqrt{X_\mathrm{m}^2+4A_0^2}}{2A_0}
\end{equation}
leading to
\begin{equation}
    F|_{P_{\mathrm{max}}}=B_0\frac{X_\mathrm{m}+\sqrt{X_\mathrm{m}^2+4A_0^2}+2A_0}{X_\mathrm{m}+\sqrt{X_\mathrm{m}^2+4A_0^2}-2A_0}\,.
\label{eq:Fano}
\end{equation}
This curve tends to the minimum Fano factor $\min_{T_H,T_C}F|_{P_\mathrm{max}}=B_0$ when $X_\mathrm{m}\gg 1$. Instead, when $X_\mathrm{m}\to 0$ it behaves as $4A_0B_0/X_\mathrm{m}\approx1.87/X_\mathrm{m}$, lying above the classical TUR bound. The comparison between this result and the minimum Fano factor as predicted from our TUR is shown in Fig.~\ref{fig:Fano_pmax}.

Regarding the efficiency, Ref.~\cite{Whitney2015Mar} obtained
\begin{equation}
    \left.\frac{\eta}{\eta_\mathrm{c}}\right|_{P_\mathrm{max}}=\left[1+C_0\left(1+\frac{T_C}{T_H}\right)\right]^{-1}\,,
\end{equation}
with $C_0=\mathrm{Li}_2(-e^{-\delta_0})/[\delta_0\ln(1+e^{-\delta_0})]\approx 0.934$. This efficiency ranges between $0.35\eta_\mathrm{c}$, when $T_C=T_H$, and $0.52\eta_\mathrm{c}$, when $T_C\ll T_H$.

%%%%%%%%%%%%%%%%%%%%%%%%%%%%%%%%%%%%%%%%%%%%%%%%%%%%%%
%%%%%%%%%%%%%%%%%%%%%%%%%%%%%%%%%%%%%%%%%%%%%%%%%%%%%

\section{Proof of the TUR bound for bosonic transport}
\label{app:bosonic}

The preceding sections considered fermionic reservoirs. We now repeat the analysis
for bosonic transport. The affinities $a_\alpha(E)=(E-\mu_\alpha)/T_\alpha$, the
variables $x(E)$, $u(E)$, the slope $\tau=(T_H+T_C)/(T_H-T_C)$, and the
energy-independent intercept $u^*$ in $u(E)=u^*+\tau x(E)$ do not involve the
statistics and are used unchanged from Sec.~\ref{appB}. Only the distribution
function and the structure of the current noise differ.

The starting point are the expressions of current and noise in terms of the variables $x(E)$ and $u(E)$, which we recall here:
\begin{subequations}
    \begin{align}
        I&=\int \frac{dE}{h}\,\mathcal{T}(E)g_I(x(E),u(E))\,,\\
        S=S_I^\mathrm{cl}+ S_I^\mathrm{qu}&=\int \frac{dE}{h}\,\mathcal{T}(E)g_S^{\rm cl}(x(E),u(E))+\int \frac{dE}{h}\,\mathcal{T}^2(E)g_I^2(x(E),u(E))\,.
    \end{align}
\end{subequations}

For single-energy transport occurring at $E=E_0$, the Fano factor obeys
\begin{equation}
    F=\frac{S_I}{|I|}\ge\frac{S_{I}^\mathrm{cl}}{|I|}=\frac{g_S^{\rm cl}(x(E_0),u(E_0))}{|g_I(x(E_0),u(E_0))|}=\coth(|x(E_0)|)=\coth X\,.
\label{eq:cross_identity}
\end{equation}
We see that the ratio $g_S^{\rm cl}/|g_I|$ is independent of $u$, meaning that it carries no information on the working mode, which is encoded entirely in
$u$, via $u^*$ (see Sec.~\ref{appB}). Therefore, \eqref{eq:cross_identity} is mode-independent.

Extending the single-energy problem to a general transmission function simply relies on the convexity of $\coth z$ for $z>0$, allowing us to use Jensen's inequality to bound the Fano factor.
Explicitly:
\begin{align}
    F&=\frac{S_I}{|I|}\ge\frac{S_I^\mathrm{cl}}{|I|}=\frac{\int dE\,\mathcal{T}(E)g_S^{\rm cl}(x(E),u(E))}{\left|\int dE\,\mathcal{T}(E)g_I(x(E),u(E))\right|}\overset{\eqref{eq:cross_identity}}{=}\frac{\int dE\,\mathcal{T}(E)|g_I(x(E),u(E))|\coth(|x(E)|)}{\left|\int dE\,\mathcal{T}(E)g_I(x(E),u(E))\right|}\nonumber\\
    &\ge\frac{\int dE\,\mathcal{T}(E)|g_I(x(E),u(E))|\coth(|x(E)|)}{\int dE\,\mathcal{T}(E)|g_I(x(E),u(E))|}\overset{\mathrm{Jensen}}{\ge}\coth\left(\frac{\int dE\,\mathcal{T}(E)|g_I(x(E),u(E))|\,|x(E)|}{\int dE\,\mathcal{T}(E)|g_I(x(E),u(E))|}\right)\nonumber\\
    &\ge\coth\left(\frac{\int dE\,\mathcal{T}(E)|g_I(x(E),u(E))|\,|x(E)|}{\left|\int dE\,\mathcal{T}(E)g_I(x(E),u(E))\right|}\right)=\coth\left(\frac{\sigma}{2|I|}\right)=\coth X\,,
\label{eq:bosonic_TUR_charge}
\end{align}
where in the second-to last line we used that $\coth z$ is a decreasing function for $z>0$, and in the last line we recognised the entropy production at the numerator (absolute values can be removed since $g_I$ and $x$ have the same sign).
The bound holds for any $0\le\mathcal{T}(E)\le1$ and in every working mode.
Since
$\coth X>1/X$, it is strictly stronger than the classical TUR, reducing to it as $X\to 0$.
We notice that this result can also be obtained by combining Eqs.~(56) and (57) of Ref.~\cite{Palmqvist2025Jul}, which are also valid for nonthermal occupations of the reservoirs.

We emphasize that having a mode-independent equality $g_S^{\rm cl}/|g_I|=\coth(|x(E)|)$, rather than a mode-dependent inequality as in the fermionic case, is what allows us to avoid the tangent-line construction here and directly use Jensen's inequality.

\section{Proof of the TUR bound for classical carriers}
\label{app:classical}
For ballistic transport of classical particles, the current, noise, and entropy production turn out to be described by very similar expressions compared to those of scattering theory~\cite{Brandner2018Mar}:
\begin{align}
    I&=\int \frac{dE}{h} \mathcal{T}(E)[n_H(E)-n_C(E)]\,,\\
    S_I&=\int \frac{dE}{h}\mathcal{T}(E)[n_H(E)+n_C(E)]\,,\\
    \sigma&=\int \frac{dE}{h}\mathcal{T}(E)[a_C(E)-a_H(E)][n_H(E)-n_C(E)]\,,
\end{align}
where $n_\alpha(E)=\exp[-a_\alpha(E)]$ are simply Maxwell-Boltzmann distributions. Then, using the same variables $x(E)$ and $u(E)$ as in the previous derivations, the functions $g_I$ and $g_S^{\rm th}$ take the simpler forms
\begin{align}
    g_I(x(E),u(E))&=2e^{-u(E)}\sinh[x(E)]\,,\\
    g_S^{\rm th}(x(E),u(E))&=2e^{-u(E)}\cosh[x(E)]\,.
\end{align}
So, we see that the structure of the ratio $g_S^{\rm th}/|g_I|$ is identical to that of the bosonic case, allowing us to repeat essentially the same calculation of Sec.~\ref{app:bosonic}. This leads to
\begin{equation}
    \frac{S_I}{|I|}\ge\coth\left(\frac{\sigma}{2|I|}\right)\,,
\end{equation}
irrespective of the working mode. Equality is \emph{always} reached for single-energy transport.

%%%%%%%%%%%%%%%%%%%%%%%%%%%%%%%%%%%%%%%%%%%%%%%%%%%%%%%%%%%%%%%%%%%%%%%%%%%%%%%%%%%%%%%%%%%%%%%%%%%%%%%%%%%%%%%%%%%%%%%%%%%%%%%%%%%%%%%%%%%%

\section{Bounds for the efficiency of a coherent thermoelectric heat engine}
\label{app:efficiency}
In this section we rewrite the different charge-current fluctuation bounds as
upper bounds on the efficiency of a two-terminal heat engine $(Q_H>0,Q_C<0,P>0)$.
The efficiency and Carnot's efficiency are
\begin{align}
    \eta=\frac{P}{Q_H},
    \qquad
    \eta_\mathrm{c}=1-\frac{T_C}{T_H}.
\end{align}
The rate of entropy production can then be written as
\begin{align}
    \sigma=-\frac{Q_H}{T_H}-\frac{Q_C}{T_C}
    =\frac{P}{T_C}\left(\frac{\eta_\mathrm{c}}{\eta}-1\right).
\end{align}
We have $P=(\mu_C-\mu_H)I=\delta\mu\, I=|\delta\mu||I|$, where the last step follows from $P>0$. Therefore,
\begin{align}
    \frac{\sigma}{|I|}  = \frac{|\delta\mu|}{T_C} \left( \frac{\eta_\mathrm{c}}{\eta}-1 \right).
\end{align}

Any lower bound on $X=\sigma/2|I|$ of the form $X\ge \Xi(F)/2$ implies
\begin{align}
    \frac{\sigma}{|I|} \ge \Xi(F).
\end{align}
Thus, all efficiency bounds descending from a TUR can be written in the compact form
\begin{align}
    \frac{\eta}{\eta_\mathrm{c}} \le \frac{1}{ 1+\frac{T_C}{|\delta\mu|}\Xi(F) }
\end{align}
For the classical TUR, one obtains
\begin{align}
    \Xi_{\rm cl}(F)=\frac{2}{F}.
\end{align}
For the Brandner-Saito TUR, $F\sinh X\ge 1$, one finds
\begin{align}
    \Xi_{\rm BS}(F)
    = 2\sinh^{-1}\!\left(\frac{1}{F}\right).
\end{align}
Finally, the heat-engine quantum bounds derived in the main text read

\begin{align}
F\ge
\begin{cases}
\dfrac{\cosh^2 X+1}{\sinh(2X)}>\frac{1}{2},
& \text{fermions},\\
\quad\coth X>1,
& \text{bosons}.
\end{cases}
\end{align}

Inverting this relation gives
$X
\ge
\frac{1}{2}\Xi_{\rm HE}(F)$,
where 
\begin{align}
\Xi_{\rm HE}(F)
=
\frac{1}{2}
\begin{cases}
\ln\left(
\dfrac{3+2\sqrt{F^2+2}}{2F-1}
\right),
& \text{fermions},\\
\ln\left(
\dfrac{F+1}{F-1}
\right),
& \text{bosons}.
\end{cases}
\label{eq:Xi_HE_cases}
\end{align}

The corresponding heat-engine efficiency bound is
\begin{align}
\frac{\eta}{\eta_\mathrm{c}}
\le
\begin{cases}
\left[
1+
\dfrac{k_{\mathrm B}T_C}{|\delta\mu|}
\ln\left(
\dfrac{3+2\sqrt{F^2+2}}{2F-1}
\right)
\right]^{-1},
& \text{fermions},\\
\left[
1+
\dfrac{k_{\mathrm B}T_C}{|\delta\mu|}
\ln\left(
\dfrac{F+1}{F-1}
\right)
\right]^{-1},
& \text{bosons}.
\end{cases}
\label{eq:eff_HE_F_B_app}
\end{align}
 Alternatively, one can express the Fano factor in terms of the output power and its fluctuation, so for a fermionic heat engine one has
\begin{figure}[t]
	\centering
	\includegraphics[width=0.9\linewidth]{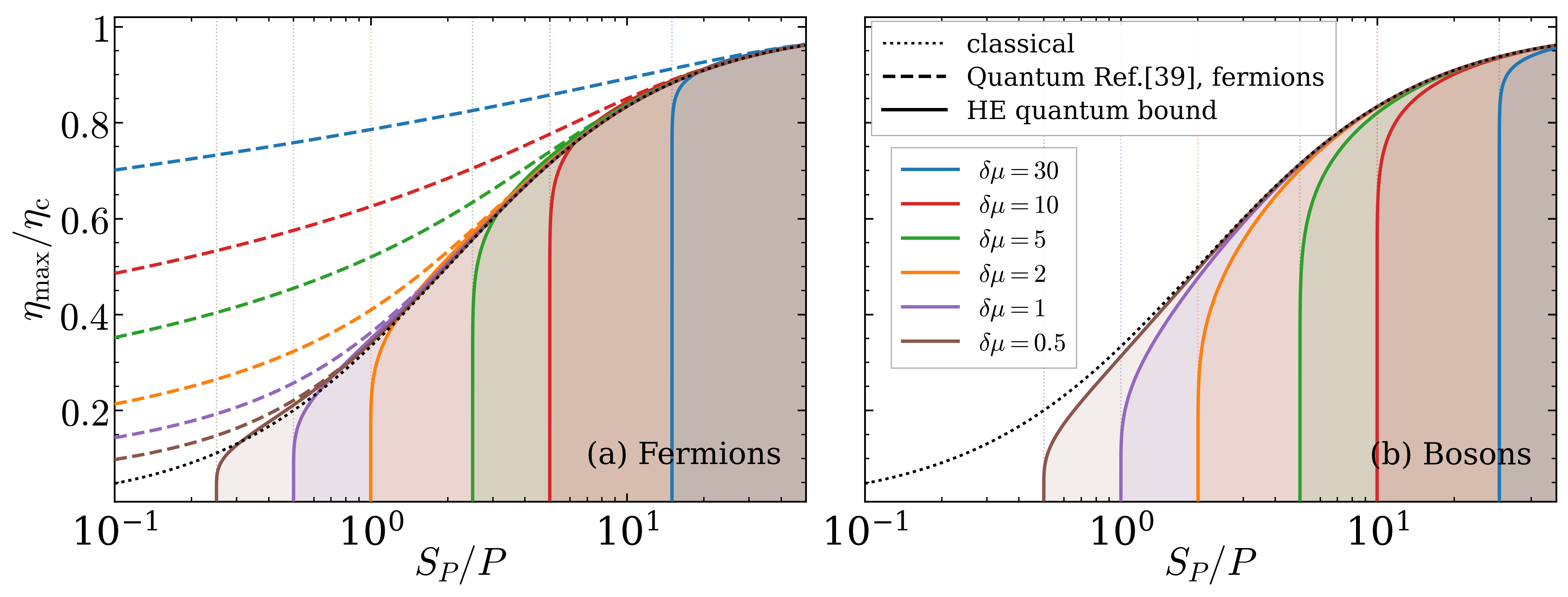}
	\caption{Maximum efficiency $\eta_\mathrm{max}/\eta_{\mathrm{C}}$ allowed by different current fluctuation bounds as a function of the output power noise to current ratio $S_{P}/P$ for (a) fermions, (b) bosons. The dotted, dashed, and solid lines correspond to the classical TUR, Brandner-Saito's quantum TUR, and the heat engine quantum bound derived here. Different colors indicate different values of  $\delta\mu$ at fixed $T_C=1$. The shaded regions correspond to the allowed parameter space by the HE bound. }
	\label{figS4}
\end{figure}
\begin{align}
    \frac{S_P}{P}>\frac{1}{2\delta\mu}
\end{align}

In \cref{figS4} we show the maximum efficiency of a heat engine as a function of $ \frac{S_P}{P}$ for different values of $\delta\mu$ and fixed $T_C=1$ for (a) fermions, and (b) bosons. Each working condition has a minimum possible value of $\frac{S_P}{P}$ to operate as a heat engine.

%%%%%%%%%%%%%%%%%%%%%%%%%%%%%%%%%%%%%%%%%%%%%%%%%%%%%%%%%%%%%%%%%%%%%%%%%%%%%%%%%%%%%%%%%%%%%%%%%%%%%%%%%%%%%%%%%%%%%%%%%%%%%%%%%%%%%%%%%%%%

\section{Multiterminal heat engines}
\label{app:multiterminal}
In this section we show that the fermionic heat-engine bound, Eq.~\eqref{eq:HE_TUR_app}, extends to multiterminal conductors, and we identify the currents to which it applies.

\begin{figure}[t]
	\centering
	\includegraphics[width=0.5\linewidth]{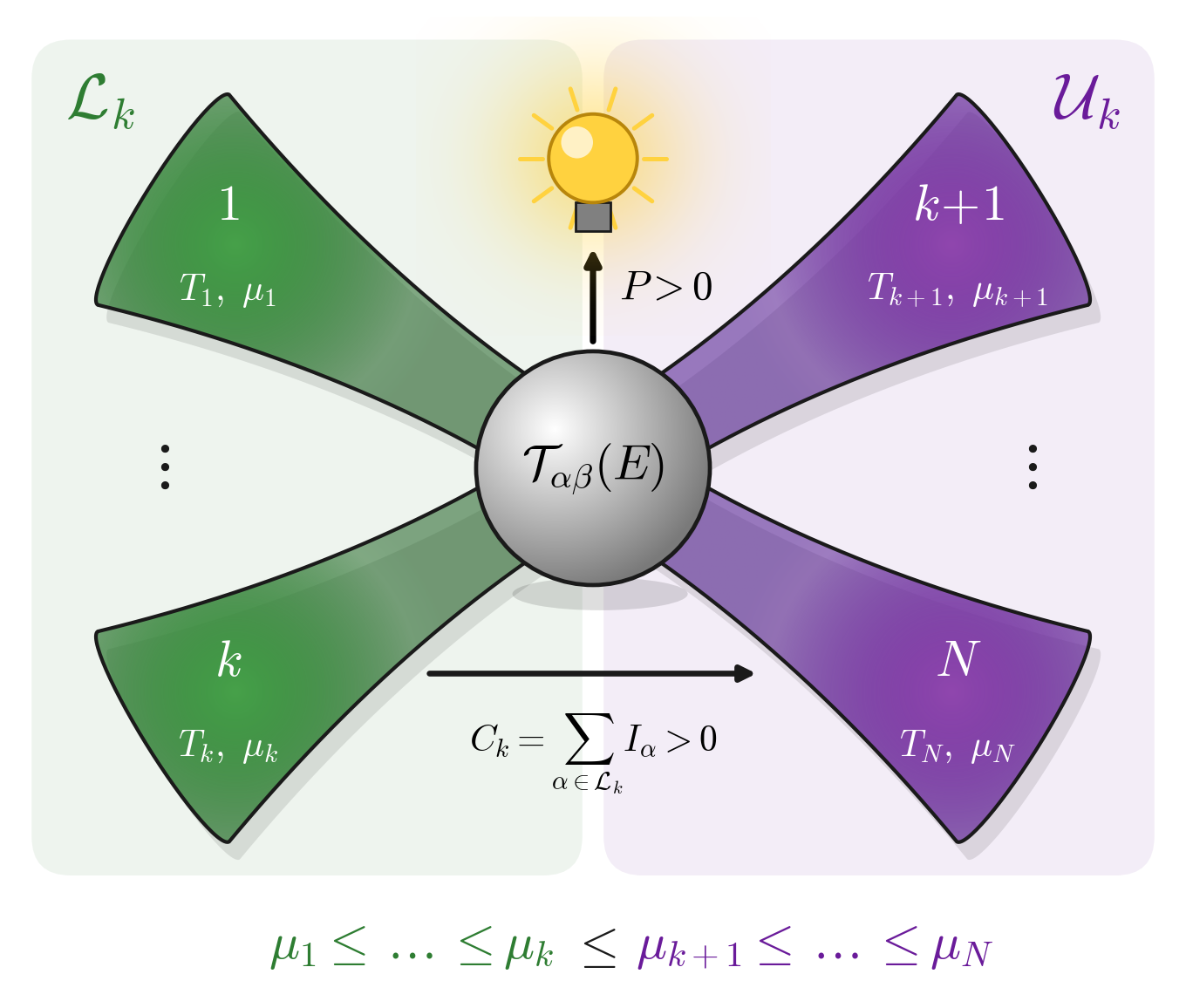}
\caption{Multiterminal coherent conductor with transmissions $\mathcal T_{\alpha\beta}(E)$, operating as a heat engine ($P>0$). The reservoirs, labeled by increasing chemical potential, $\mu_1\le\mu_2\le\dots\le\mu_N$, are partitioned into the set $\mathcal{L}_k=\{1,\dots,k\}$ (green) and its complement $\mathcal{U}_k=\{k+1,\dots,N\}$ (purple). The temperatures $T_\alpha$ are arbitrary. The arrow represents the net particle current across the partition, $C_\mathcal{L}=\sum_{\alpha\in \mathcal{L}}I_\alpha$, which flows against the chemical-potential difference when $C_\mathcal{L}>0$.}
\label{figS5}
\end{figure}

We consider $N$ fermionic reservoirs with temperatures $T_\alpha$, chemical potentials $\mu_\alpha$, affinities $a_\alpha(E)=(E-\mu_\alpha)/T_\alpha$, and occupations $n_\alpha(E)=[e^{a_\alpha(E)}+1]^{-1}$, connected by a coherent conductor with scattering matrix $s(E)$, as illustrated in \cref{figS5}. Lead $\alpha$ supports $N_\alpha$ channels, and $s_{\alpha\beta}$ denotes the $N_\alpha\times N_\beta$ block describing scattering from lead $\beta$ into lead $\alpha$. The transmissions are $\mathcal T_{\alpha\beta}(E)=\Tr[s^\dagger_{\alpha\beta}s_{\alpha\beta}]$. Unitarity of $s$, $ss^\dagger=s^\dagger s=\mathds 1$, reads in blocks
\begin{equation}
\sum_\gamma s_{\alpha\gamma}s^\dagger_{\beta\gamma}=\delta_{\alpha\beta}\mathds 1_{N_\alpha},\qquad
\sum_\alpha s^\dagger_{\alpha\gamma}s_{\alpha\delta}=\delta_{\gamma\delta}\mathds 1_{N_\gamma},
\label{eq:mt_unitarity}
\end{equation}
and, taking traces, $\sum_\beta\mathcal T_{\alpha\beta}=N_\alpha$. We assume time-reversal symmetry (TRS), $\mathcal T_{\alpha\beta}=\mathcal T_{\beta\alpha}$. $I_\alpha$ and $Q_\alpha$ denote the particle and heat currents flowing out of reservoir $\alpha$, so that $\sum_\alpha I_\alpha=0$, and
\begin{equation}
\sigma=-\sum_\alpha\frac{Q_\alpha}{T_\alpha},\qquad P=\sum_\alpha Q_\alpha=-\sum_\alpha\mu_\alpha I_\alpha .
\end{equation}

For every pair of reservoirs we introduce the variables
\begin{equation}
x_{\alpha\beta}(E)=\frac{a_\beta(E)-a_\alpha(E)}{2},\qquad u_{\alpha\beta}(E)=\frac{a_\alpha(E)+a_\beta(E)}{2},
\end{equation}
which generalize $x(E)$ and $u(E)$ of Sec.~\ref{appA}, so that, with the functions of Eq.~\eqref{eq:gI_gS_y},
\begin{equation}
n_\alpha-n_\beta=g_I(x_{\alpha\beta},u_{\alpha\beta}),\qquad
n_\alpha(1-n_\alpha)+n_\beta(1-n_\beta)=g_S^{\rm th}(x_{\alpha\beta},u_{\alpha\beta}).
\label{eq:mt_pair_functions}
\end{equation}
Note that $x\,g_I(x,u)\ge0$. The particle and heat currents leaving reservoir $\alpha$ are
\begin{equation}
I_\alpha=\sum_{\beta\neq\alpha}\int\frac{dE}{h}\,\mathcal T_{\alpha\beta}\,g_I(x_{\alpha\beta},u_{\alpha\beta}),\qquad
Q_\alpha=\sum_{\beta\neq\alpha}\int\frac{dE}{h}\,(E-\mu_\alpha)\,\mathcal T_{\alpha\beta}\,g_I(x_{\alpha\beta},u_{\alpha\beta}),
\label{eq:mt_I}
\end{equation}
and, symmetrizing in $\alpha\leftrightarrow\beta$ with TRS, the entropy production can be written as~\cite{brandner2025thermodynamic}
\begin{equation}
\sigma=\frac12\int\frac{dE}{h}\sum_{\alpha,\beta}\mathcal T_{\alpha\beta}\,(a_\beta-a_\alpha)(n_\alpha-n_\beta)
=\sum_{\alpha<\beta}\sigma_{\alpha\beta},\qquad
\sigma_{\alpha\beta}=2\int\frac{dE}{h}\,\mathcal T_{\alpha\beta}\,x_{\alpha\beta}\,g_I(x_{\alpha\beta},u_{\alpha\beta})\ge0 .
\label{eq:mt_sigma}
\end{equation}
The mean particle and heat currents, and hence $\sigma$ and $P$, are thus sums of two-terminal contributions (``pairs''), and each $\sigma_{\alpha\beta}$ is non-negative.

We partition the reservoirs into a set $\mathcal{L}$ and its complement $\mathcal{U}$, and define the cut current
\begin{equation}
C_\mathcal{L}=\sum_{\alpha\in \mathcal{L}}I_\alpha ,
\end{equation}
i.e., the net particle current flowing from $\mathcal{L}$ into $\mathcal{U}$. Relevant partitions for the heat-engine bound, in which all reservoirs of $\mathcal{L}$ have lower chemical potentials than those of $\mathcal{U}$, is sketched in Fig.~\ref{figS5} (where $\mathcal{L}=\mathcal{L}_k$ and $\mathcal{U}=\mathcal{U}_k$).
When summing Eq.~\eqref{eq:mt_I} over $\alpha$ to obtain $C_\mathcal{L}$, pairs with both reservoirs in $\mathcal{L}$ contribute $\mathcal T_{\alpha\beta}(n_\alpha-n_\beta)+\mathcal T_{\beta\alpha}(n_\beta-n_\alpha)=0$ by TRS, so only pairs crossing the partition survive:
\begin{equation}
C_\mathcal{L}=\sum_{\alpha\in \mathcal{L}}\sum_{\beta\in \mathcal{U}}\int\frac{dE}{h}\,\mathcal T_{\alpha\beta}\,g_I(x_{\alpha\beta},u_{\alpha\beta}).
\label{eq:mt_CL}
\end{equation}
Correspondingly, we define the entropy production of all crossing pairs,
\begin{equation}
\sigma_{\mathcal{LU}}=\sum_{\alpha\in \mathcal{L}}\sum_{\beta\in \mathcal{U}}2\int\frac{dE}{h}\,\mathcal T_{\alpha\beta}\,x_{\alpha\beta}\,g_I(x_{\alpha\beta},u_{\alpha\beta})\ \le\ \sigma ,
\label{eq:mt_sigmaLU}
\end{equation}
where the inequality follows from the non-negativity of all $\sigma_{\alpha\beta}$ in Eq.~\eqref{eq:mt_sigma}.

The zero-frequency noise of $C_\mathcal{L}$ is $S_{C_\mathcal{L}}=\sum_{i,j\in \mathcal{L}}S_{ij}$, and therefore contains the cross-correlations between pairs of currents $(I_i,I_j)$ in terminals belonging to $\mathcal{L}$. The elements of the noise matrix are given by~\cite{Blanter2000Sep}
\begin{equation}
S_{ij}=\frac12\int\frac{dE}{h}\sum_{\alpha,\beta}\Tr\big[A_{\alpha\beta}(i)A_{\beta\alpha}(j)\big]\big\{n_\alpha(1-n_\beta)+n_\beta(1-n_\alpha)\big\},\qquad
A_{\alpha\beta}(i)=\delta_{i\alpha}\delta_{i\beta}\,\mathds 1-s^\dagger_{i\alpha}s_{i\beta},
\label{eq:mt_Buttiker}
\end{equation}
with $s^\dagger_{i\alpha}\equiv(s_{i\alpha})^\dagger$ and $\mathds 1$ the $N_i\times N_i$ identity. Summing over $i,j\in \mathcal{L}$, and using that the trace is linear in each of its two factors,
\begin{equation}
S_{C_\mathcal{L}}=\frac12\int\frac{dE}{h}\sum_{\alpha,\beta}W_{\alpha\beta}\big\{n_\alpha(1-n_\beta)+n_\beta(1-n_\alpha)\big\},\qquad
W_{\alpha\beta}\equiv\Tr\Big[\Big(\sum_{i\in \mathcal{L}}A_{\alpha\beta}(i)\Big)\Big(\sum_{j\in \mathcal{L}}A_{\beta\alpha}(j)\Big)\Big],
\label{eq:mt_SL_W}
\end{equation}
where
\begin{equation}
\sum_{i\in \mathcal{L}}A_{\alpha\beta}(i)=
\begin{cases}
\delta_{\alpha\beta}\mathds 1-\sum_{i\in \mathcal{L}}s^\dagger_{i\alpha}s_{i\beta}, & \alpha\in \mathcal{L},\\[3pt]
-\sum_{i\in \mathcal{L}}s^\dagger_{i\alpha}s_{i\beta}, & \alpha\in \mathcal{U}.
\end{cases}
\label{eq:mt_sumA}
\end{equation}
Since $(s^\dagger_{i\alpha}s_{i\beta})^\dagger=s^\dagger_{i\beta}s_{i\alpha}$, the second factor in $W_{\alpha\beta}$ is the Hermitian conjugate of the first, so that $W_{\alpha\beta}=\Tr[XX^\dagger]\ge0$, and by cyclicity of the trace $W_{\alpha\beta}=W_{\beta\alpha}$. Inserting $n_\alpha(1-n_\beta)+n_\beta(1-n_\alpha)=n_\alpha(1-n_\alpha)+n_\beta(1-n_\beta)+(n_\alpha-n_\beta)^2$ into Eq.~\eqref{eq:mt_SL_W} and using the symmetry of $W$,
\begin{equation}
S_{C_\mathcal{L}}=\int\frac{dE}{h}\sum_{\alpha,\beta} W_{\alpha\beta}\,n_\alpha(1-n_\alpha)
+\frac12\int\frac{dE}{h}\sum_{\alpha,\beta}W_{\alpha\beta}(n_\alpha-n_\beta)^2\equiv S_{C_\mathcal{L}}^{\mathrm{th}}+S_{C_\mathcal{L}}^{\mathrm{sh}},
\label{eq:mt_split}
\end{equation}
where $S_{C_\mathcal{L}}^{\rm sh}\ge0$, since each of its terms is a product of non-negative factors.

To evaluate $S_\mathcal{L}^{\rm th}$ we need $\sum_\beta W_{\alpha\beta}$. For $\alpha\in \mathcal{L}$, inserting the first line of Eq.~\eqref{eq:mt_sumA} into both factors gives
\begin{equation}
\sum_\beta W_{\alpha\beta}=N_\alpha-2\sum_{i\in \mathcal{L}}\mathcal T_{i\alpha}+\Tr\Big[\sum_\beta\Big(\sum_{i\in \mathcal{L}}s^\dagger_{i\alpha}s_{i\beta}\Big)\Big(\sum_{j\in \mathcal{L}}s^\dagger_{j\beta}s_{j\alpha}\Big)\Big].
\end{equation}
Using the first relation in Eq.~\eqref{eq:mt_unitarity}, $\sum_\beta s_{i\beta}s^\dagger_{j\beta}=\delta_{ij}\mathds 1$, the last term is
\begin{equation}
\sum_{i,j\in \mathcal{L}}\Tr\Big[s^\dagger_{i\alpha}\Big(\sum_\beta s_{i\beta}s^\dagger_{j\beta}\Big)s_{j\alpha}\Big]=\sum_{i\in \mathcal{L}}\mathcal T_{i\alpha},
\label{eq:mt_unit}
\end{equation}
so that, with $\sum_i\mathcal T_{i\alpha}=N_\alpha$, $\sum_\beta W_{\alpha\beta}=N_\alpha-\sum_{i\in \mathcal{L}}\mathcal T_{i\alpha}=\sum_{\beta\in \mathcal{U}}\mathcal T_{\beta\alpha}$. For $\alpha\in \mathcal{U}$, the second line of Eq.~\eqref{eq:mt_sumA} has no $\delta$ term, and Eq.~\eqref{eq:mt_unit} directly gives $\sum_\beta W_{\alpha\beta}=\sum_{i\in \mathcal{L}}\mathcal T_{i\alpha}$. Renaming the indices so that $\alpha\in \mathcal{L}$ and $\beta\in \mathcal{U}$, and using TRS, we obtain
\begin{equation}
S_{C_\mathcal{L}}\ \ge\ S_{C_\mathcal{L}}^{\rm th}=\sum_{\alpha\in \mathcal{L}}\sum_{\beta\in \mathcal{U}}\int\frac{dE}{h}\big[\mathcal T_{\beta\alpha}n_\alpha(1-n_\alpha)+\mathcal T_{\alpha\beta}n_\beta(1-n_\beta)\big]
py=\sum_{\alpha\in \mathcal{L}}\sum_{\beta\in \mathcal{U}}\int\frac{dE}{h}\,\mathcal T_{\alpha\beta}\,g_S^{\rm th}(x_{\alpha\beta},u_{\alpha\beta}),
\label{eq:mt_SL}
\end{equation}
which generalizes $S_I\ge S_I^{\rm th}$ to multiterminal conductors. Pairs of reservoirs on the same side of the partition do not appear: a particle transferred between two reservoirs of $\mathcal{L}$ does not change $C_\mathcal{L}$. Equations~\eqref{eq:mt_CL}, \eqref{eq:mt_sigmaLU}, and~\eqref{eq:mt_SL} show that, with respect to the partition $(\mathcal{L},\mathcal{U})$, a multiterminal conductor has exactly the structure of a two-terminal one: the current, the entropy production of the crossing pairs, and the lower bound on the noise are the same integrals as in Eq.~\eqref{eq:I_sigma_S}, with $\int dE\,\mathcal T(E)$ replaced by $\sum_{\alpha\in \mathcal{L},\beta\in \mathcal{U}}\int dE\,\mathcal T_{\alpha\beta}(E)$.

\subsection{Pointwise inequality for a pair of reservoirs}
The proof of Sec.~\ref{appB} relies on the pointwise inequality~\eqref{eq:ineq_gS_lXgI}, $g_S^{\rm th}(x,u)\ge l_X(x)\,g_I(x,u)$, with $l_X$ the tangent line to $y^{\min}_{\rm EF}$ at an arbitrary $X>0$. Since $y(x,u)$ increases with $|u|$ and equals $y^{\min}_{\rm EF}(x)$ at $|u|=x$, the inequality holds at all points with $x>0$ and $|u|\ge x$, exactly as in case (i) of Sec.~\ref{appB}; for $x<0$ it holds automatically, as in case (ii). It can therefore fail only at points with $x>0$ and $|u|<x$.

For a pair with $x_{\alpha\beta}>0$, i.e., $a_\beta>a_\alpha$, the condition $|u_{\alpha\beta}|\ge x_{\alpha\beta}$ is equivalent to $|a_\alpha+a_\beta|\ge a_\beta-a_\alpha$, which fails only if $a_\alpha<0<a_\beta$. Since temperatures are positive, the signs of the $a_\alpha$ variables are fixed by the chemical potentials alone, and the excluded points are those in the energy window $\mu_\beta<E<\mu_\alpha$. In particular, if $\mu_\beta\ge\mu_\alpha$, Eq.~\eqref{eq:ineq_gS_lXgI} holds for the pair $(\alpha,\beta)$ at all energies, for any choice of $X>0$ and any temperatures.

\subsection{Bound for arbitrary multiterminal heat engines}
We label the reservoirs by increasing chemical potential, $\mu_1\le\mu_2\le\dots\le\mu_N$, and consider the partitions
\begin{equation}
\mathcal{L}_k=\{1,\dots,k\},\qquad \mathcal{U}_k=\{k+1,\dots,N\},\qquad k=1,\dots,N-1 ,
\end{equation}
with cut currents $C_{k}=\sum_{\alpha=1}^kI_\alpha$. With the steps of the chemical-potential ladder, $\Delta_k=\mu_{k+1}-\mu_k\ge0$, and using $\sum_\alpha I_\alpha=0$, the output power can be written as
\begin{equation}
P=-\sum_{\alpha=1}^N\mu_\alpha I_\alpha=\sum_{\alpha=1}^N(\mu_N-\mu_\alpha)\,I_\alpha=\sum_{\alpha=1}^{N}\sum_{k=\alpha}^{N-1}\Delta_k\,I_\alpha=\sum_{k=1}^{N-1}\Delta_k\sum_{\alpha=1}^{k}I_\alpha=\sum_{k=1}^{N-1}\Delta_k\,C_{k} ,
\label{eq:mt_P_cuts}
\end{equation}
where in the fourth equality we exchanged the order of the two sums, which run over all pairs $(\alpha,k)$ with $1\le\alpha\le k\le N-1$. Since all $\Delta_k\ge0$, if all $C_{k}\le0$ then $P\le0$. Hence, for a heat engine, $P>0$ implies $C_{k}>0$ for at least one $k$, i.e., particles flow on average from $\mathcal{L}_k$ to $\mathcal{U}_k$, against the chemical-potential difference.

Consider any $k$ with $C_{k}>0$. At least one crossing pair then carries a nonzero current, so $\sigma_{\mathcal{L}_k\mathcal{U}_k}>0$, and we can set $X=\sigma_{\mathcal{L}_k\mathcal{U}_k}/2C_{k}>0$. For every crossing pair, $\alpha\in \mathcal{L}_k$ and $\beta\in \mathcal{U}_k$, we have $\mu_\beta\ge\mu_\alpha$, so the window $\mu_\beta<E<\mu_\alpha$ is empty and Eq.~\eqref{eq:ineq_gS_lXgI} holds at every energy. Multiplying it by $\mathcal T_{\alpha\beta}(E)\ge0$, integrating, summing over the crossing pairs, and using Eqs.~\eqref{eq:mt_CL}, \eqref{eq:mt_sigmaLU}, and~\eqref{eq:mt_SL},
\begin{equation}
S_{C_k}\ \ge\ S^{\rm th}_{C_k}\ \ge\ \sum_{\alpha\in \mathcal{L}_k}\sum_{\beta\in \mathcal{U}_k}\int\frac{dE}{h}\,\mathcal T_{\alpha\beta}\,\big[p(X)+m(X)\,x_{\alpha\beta}\big]\,g_I(x_{\alpha\beta},u_{\alpha\beta})
=p(X)\,C_{k}+m(X)\,\frac{\sigma_{\mathcal{L}_k\mathcal{U}_k}}{2}.
\end{equation}
Dividing by $C_{k}>0$ gives $S_{C_k}/C_{k}\ge p(X)+m(X)X=l_X(X)=y^{\min}_{\rm EF}(X)$. Finally, since $\sigma\ge\sigma_{\mathcal{L}_k\mathcal{U}_k}$ and $y^{\min}_{\rm EF}$ is decreasing,
\begin{equation}
\frac{S_{C_k}}{C_{k}}\ \ge\ y^{\min}_{\rm EF}\!\left(\frac{\sigma_{\mathcal{L}_k\mathcal{U}_k}}{2C_{k}}\right)\ \ge\ y^{\min}_{\rm EF}\!\left(\frac{\sigma}{2C_{k}}\right)\ >\ \frac12 .
\label{eq:mt_bound}
\end{equation}

Inverting $F=y^{\min}_{\rm EF}(X)$, which yields $X=\tfrac12\ln[(3+2\sqrt{F^2+2})/(2F-1)]$, the bound can be equivalently written as a lower bound on the entropy production in terms of the Fano factor $F_k=S_{C_k}/C_{k}$,
\begin{equation}
\sigma\ \ge\ B_k\equiv C_{k}\,\ln\!\left(\frac{3+2\sqrt{F_k^2+2}}{2F_k-1}\right).
\label{eq:mt_sigma_bound}
\end{equation}
This holds for every $k$ with $C_{k}>0$, so that the most informative lower bound is $\sigma\ \ge\ \max_{k:\,C_{k}>0}B_k
$. Different $k$ involve different currents and noises, which are not ordered in general, so the optimal $k$ is determined by the values of $C_{k}$ and $S_{C_k}$ of the system under consideration.

\bibliography{biblio}